\documentclass[aps,pra,english,10pt,superscriptaddress,twocolumn,float,floatfix]{revtex4-2}  
\usepackage[english]{babel}
\usepackage[utf8]{inputenc}
\usepackage[colorinlistoftodos, color=green!40, prependcaption]{todonotes}
\usepackage{float}
\usepackage{mathtools}
\usepackage{xcolor}
\usepackage{graphicx}
\usepackage[T1]{fontenc}
\usepackage{lipsum}
\usepackage{csquotes}
\usepackage{amsmath}
\usepackage{amssymb}
\usepackage{amsfonts}
\usepackage{bbm}
\usepackage[normalem]{ulem}

\usepackage[pdftex, pdftitle={Article}, pdfauthor={Author}]{hyperref} 
\usepackage{ dsfont }
\begin{document}

\title{Caustics in the sine-Gordon model from quenches in coupled 1D Bose gases}


\author{Aman Agarwal}
    \email[]{aagarw03@uoguelph.ca}
     \affiliation{Department of Physics and Astronomy, McMaster University, 1280 Main St. W., Hamilton, Ontario, Canada L8S 4M1}
    \affiliation{BITS-Pilani, K. K. Birla Goa Campus, NH17B, Bypass Road, Zuarinagar, Goa 403726, India}
    \affiliation{International Centre for Theoretical Sciences, Tata Institute of Fundamental Research, Bengaluru -- 560089, India}
    \affiliation{Perimeter Institute for Theoretical Physics, Waterloo, Ontario, Canada, N2L 2Y5}
\affiliation{Department of Physics, University of Guelph, Guelph, Ontario, Canada, N1G 2W1}
\affiliation{Institute of Physics, University of Greifswald, 17489 Greifswald, Germany}

\author{Manas Kulkarni}
    \email[]{manas.kulkarni@icts.res.in }
    \affiliation{International Centre for Theoretical Sciences, Tata Institute of Fundamental Research, Bengaluru -- 560089, India}
    
\author{D. H. J. O’Dell}
    \email[]{dodell@mcmaster.ca }
    \affiliation{Department of Physics and Astronomy, McMaster University, 1280 Main St. W., Hamilton, Ontario, Canada L8S 4M1}
\date{\today} 

\begin{abstract}
Caustics are singularities that occur naturally in optical, hydrodynamic and quantum waves, giving rise to high amplitude patterns that can be described using catastrophe theory. In this paper we study caustics in a statistical field theory setting in the form of the sine-Gordon model that describes a variety of physical systems including coupled 1D superfluids. Specifically, we use classical field simulations to study the dynamics of two ultracold 1D Bose gases (quasicondensates) that are suddenly coupled to each other and find that the resulting  non-equilibrium dynamics are dominated by caustics.   Thermal noise is included by sampling the initial states from a Boltzmann distribution for phononic excitations.  We find that caustics pile up over time in both the number and phase difference observables leading to a characteristic non-thermal `circus tent' shaped probability distribution at long times. 
\end{abstract}

\keywords{coupled condensates, sine-Gordon hamiltonian,  cigar shaped 1-D BEC, caustics}

\maketitle

  \section{Introduction} \label{sec:outline}
   Wave focusing is ubiquitous in nature and leads to localized regions of high amplitude called caustics that dominate wavefields.  Everyday examples are provided by rainbows and also the bright lines on the bottom of water pools which are caused by the focusing of sunlight by raindrops and surface water waves, respectively \cite{Nye_natural_focusing}.
   Caustics also occur in water waves themselves as ship wakes \cite{Kelvin1905}, in the vicinity of a vortex \cite{Ravichandran2015,Deepu2017,Ravichandran2022}, and more dramatically as tsunamis (focused by the topography of the seabed \cite{Titov2005,Berry2007,Degueldre2016}) and tidal bores (focused by V-shaped bays \cite{Berry2018}). Astrophysical examples include gravitational lensing by matter  and the twinkling of starlight due to time-dependent fluctuations in the density of Earth's atmosphere. Natural focusing also leads to the phenomenon of branched flow \cite{Heller21} and is speculated to have given rise to the filamented nature of the large scale structure of the universe \cite{peeblesbook,Arnold1982,Gurbatov2012,Feldbrugge2018}. In all these systems caustics give rise to extreme amplitude fluctuations that occur more frequently than those predicted by gaussian statistics \cite{Berry77}. 
    
    A remarkable property of caustics is that they commonly take on particular characteristic shapes.  This is because caustics are singularities of the ray description, i.e.\ they are places where two or more rays coalesce leading to a diverging intensity in the short wavelength limit \cite{Berry1981SingularitiesIW}. Such singularities are described by Thom's catastrophe theory which rigorously shows that only certain shapes of singularity are structurally stable against perturbations and hence occur under `natural' or generic conditions \cite{Thom75,Arnold75,Zeeman77}. These special shapes or catastrophes form a hierarchy organized by dimension where the higher ones contain the lower ones. Each member of the hierarchy represents a class of equivalent shapes that can be smoothly transformed into each other, but each class is  distinct and cannot be smoothly transformed into any of the others. In two dimensions the only structurally stable shape is the cusp and we shall see it appear frequently when we plot quantities such as number fluctuations versus time.  It is worth noting in this context that the humble point focus that we associate with lensing is structurally unstable and unfolds into an extended caustic in the presence of perturbations (aberrations). Natural lenses are of course never perfect and so typically produce the shapes predicted by catastrophe theory. The upshot of all this is that caustics represent a form of universality in \textit{nonequilibrium wave dynamics}: they fall into equivalence classes each with their own shapes and scaling properties analogous to, but a generalization of,  equilibrium phase transitions \cite{Berry1981SingularitiesIW,Mumford2017}.

  Caustics should equally be present in quantum waves where, due to the probabilistic interpretation, they correspond to regions of high probability density.  Quantum matter wave caustics  have been seen in experiments with cold neutrons \cite{neutron2002,Jenke2011}, electron microscopes \cite{Petersen2013}, atom optics  \cite{Rooijakkers2003,Huckans09,Rosenblum2014}, and most recently in atom lasers \cite{Mossman2021}. Theoretical works on such matter wave caustics have also considered their `fine structure'  \cite{Berry1981SingularitiesIW}  which features a lattice of vortices  \cite{Simula2013,Mumford2019,Kirkby2019}. 
  Quantum fields are another area where caustics are expected to form naturally during dynamics. Early work centred on the electromagnetic field \cite{Berry04,Berry_three_quantum_obsessions}, including an interpretation of Hawking radiation as a `quantum catastrophe' \cite{Leonhardt02}, and more recently this idea has been extended to quantum many-particle systems including bosonic Josephson junctions \cite{Mumford2019,DuncanSirPRL,Goldberg2019}, the XY model with long-range interactions (Hamiltonian mean field model) \cite{Plestid2018}, quantum spin chains \cite{Kirkby2019} and the Bose-Hubbard model \cite{Kirkby2022}. One point to appreciate is that the caustics in many-body systems can occur in the wavefunction associated with an entire $N$-body configuration. Quantum many-particle caustics therefore live in Fock space which can have a large number of dimensions and hence lead to very complicated catastrophes \cite{Kirkby2022}. However, catastrophes obey projection identities which means that when projected down to lower dimensions one obtains either the same catastrophe or one lower down the hierarchy \cite{Berry1980}. Thus, low order correlation functions obtained by integrating out most of the degrees of freedom will also generically contain caustics \cite{Kirkby2019}.

\begin{figure}
\includegraphics[scale=0.6]{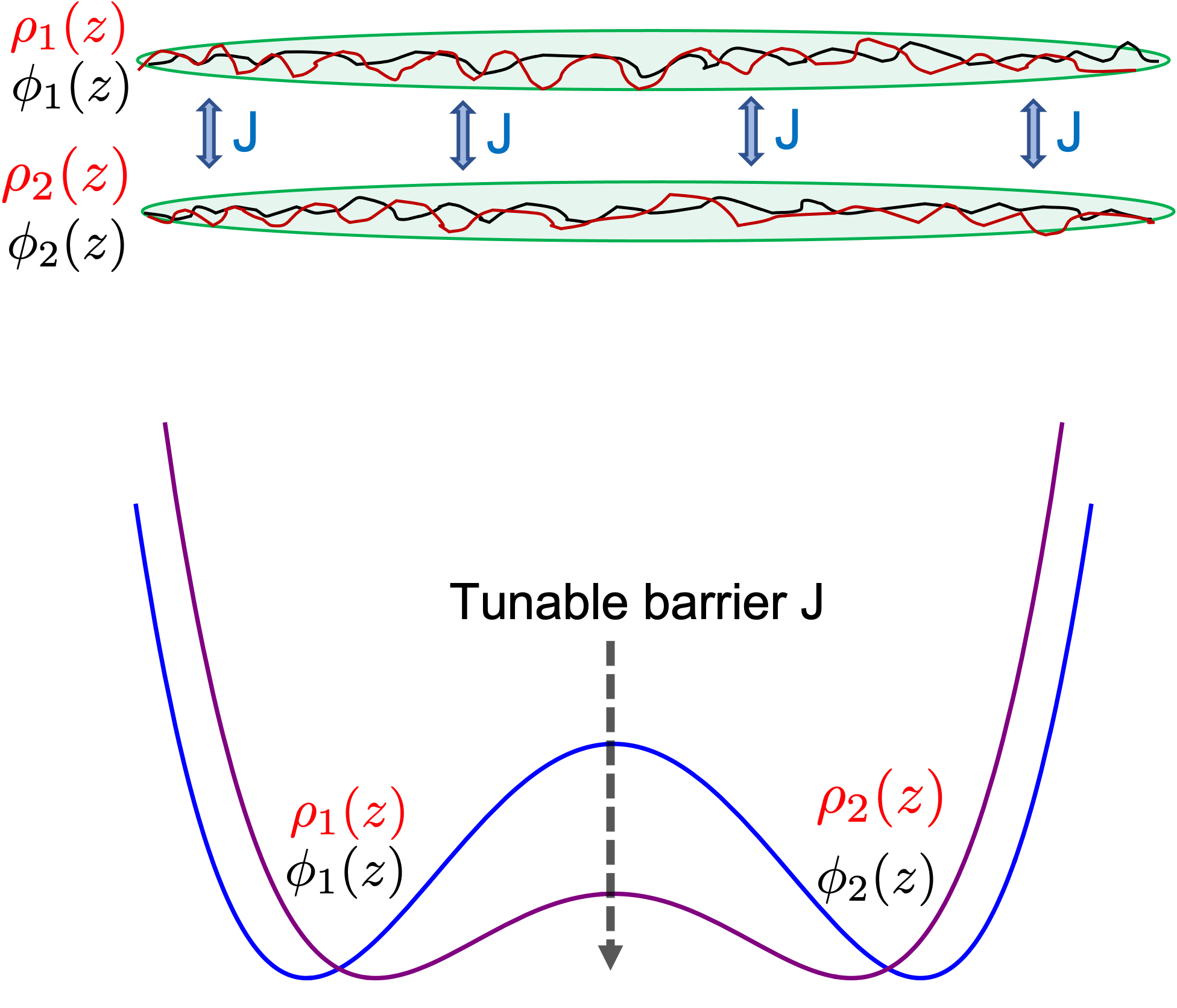}
\caption{Schematic of the setup we consider. The top figure shows two quasi one-dimensional gases that are prepared independently and then suddenly coupled together. We call this process of sudden coupling a ``$J$-quench''. $\rho_1(z)$ and $\rho_2(z)$ represent the density (red) in the first and second condensates, respectively. Similarly, $\phi_1(z)$ and $\phi_2(z)$ represent the phases (black) of the two condensates. Prior to the
$J$-quench, these fields in the two condensates are independent and contain thermal fluctuations. The bottom figure shows how a $J$-quench could be implemented by suddenly reducing the tunneling barrier height in a double well potential from a higher to a lower value.}
\label{fig:sch}
\end{figure}  

  In this paper we study caustics in the sine-Gordon (SG) model. The (classical)  SG model obeys the nonlinear wave equation
  \begin{equation}
      \frac{\partial^2 \phi}{\partial t^2}-c_{0}^2 \frac{\partial^2 \phi}{\partial z^2} +\omega_{0}^2 \sin \phi =0
      \label{eq:waveeqn}
  \end{equation}
  where $\phi=\phi(z,t)$ is a one dimensional field, and $c_{0}$ and $\omega_{0}$ represent a characteristic speed and frequency, respectively. If $c_{0}$ is taken to be the speed of light then Eq.\ (\ref{eq:waveeqn}) is relativistically covariant, being a nonlinear version of the Klein-Gordon equation and reducing to it when $\phi \ll 1$ such that $\sin \phi \approx \phi$. The SG model received attention from the high energy physics community in the 1970s due its soliton solutions  \cite{Coleman1975,Rajaraman75,Fogel1977,Sine_Gordon_book}, but also describes the low energy physics of a considerable range of condensed matter systems including crystal dislocations \cite{Frenkel1939}, domain walls in magnetic \cite{Enz1964} and binary superfluid \cite{Gallemi2019} systems, mesoscopic thin-walled superconducting cylinders in magnetic fields \cite{sg_superconductor1,sg_superconductor2}, the Heisenberg spin chain with a field induced gap \cite{Oshikawa1997,Affleck1999,Cubero2017}, trapped ions \cite{morigi2022}, two-dimensional Bose gases realizing the XY model \cite{Hadzibabic2006}, one-dimensional Bose gases in periodic potentials (that can capture the Mott-insulator to superfluid transition in one dimension) \cite{Haldane1981,Giamarchi_book}, one-dimensional ultracold bosonic gases with two spin states \cite{widera2008}, and two tunnel-coupled one-dimensional single component Bose gases \cite{Bouchoule2005,Gritsev_2007_linear_response,iucci2010,Torre2013,Foini2015,Schmiedmayer,vanNieuwkerk2021,Mennemann2021}.   
  The fact that the SG model is both nonlinear and integrable means that attention is often focused on its soliton solutions, but part of our mission in this paper is to point out that these same properties also imply that caustics (which are associated with the existence of tori in phase space \cite{Berry1983Chaos}) are expected to occur generically, and we are aware of only one previous study of caustics in this model \cite{SGcaustic1}.

  The particular physical realization we have in mind for this paper is a system composed of two elongated quasi-one-dimensional single component Bose gases coupled by tunneling along their length; the field $\phi(z,t)$ in Eq.\ (\ref{eq:waveeqn}) gives the relative phase between the two quantum gases.
  Quasi one-dimensional Bose gases have been created in a number of experiments over the last two decades using tightly trapped ultracold atoms,  and the remarkable tunability of these systems allows the strongly interacting Tonks-Girardeau regime \cite{Paredes2004,Kinoshita2004}, the weakly interacting quasicondensate regime \cite{esteve06,Hofferberth2008,Schley2013,fang16}, and also the crossover between the two \cite{Jacqmin2011,Schemmer2019}, to be reached.
  It is important to note that, in accordance with the Mermin-Wagner theorem \cite{Mermin1966}, one-dimensional Bose gases do not undergo true Bose-Einstein condensation at low temperature, unlike three dimensional gases. Instead, they can form quasicondensates where density fluctuations are still quite suppressed but phase fluctuations that destroy off-diagonal long range order remain \cite{quantum_regimes_paper,Kheruntsyan2003,Kheruntsyan2005}.  In this paper we shall work in the weakly interacting regime and assume a state of the system consisting of a  quasicondensate plus small thermal fluctuations.

  A system comprised of two coupled quasi-one dimensional gases can be made by taking a single gas and splitting it in two along its long axis by switching on an elongated double well potential. This is the experimental protocol typically adopted in a series of experiments conducted by the Vienna group \cite{Hofferberth2008,Betz2011,Gring_and_Schmeidmayer_relaxation_paper,Local_emergence_and_correlation_paper,Schweigler_2017,Schmeidmayer_double_light_cone_paper,Rauer2018,Pigneur2018}. The combination of almost complete isolation from the environment, long relaxation times and spatially resolved measurements of phase and number difference make these experiments ideal for investigating many-particle quantum dynamics, including fundamental questions such as whether and how closed quantum systems reach equilibrium.  The gas can be split slowly so that it always remains close to equilibrium leading to number squeezed states \cite{Esteve08,Berrada2013} or it can be split rapidly, leading to a so-called quantum quench which launches the system into a nonequilibrium state.

   In this paper we shall consider the \textit{opposite} quench where two one-dimensional gases are suddenly connected together (see schematic representation in Figure \ref{fig:sch}). This touches on rather fundamental considerations in quantum mechanics since it describes the build-up of coherence between two initially independent systems, and is therefore related to the double-slit experiment for many-particle systems \cite{andrews97,castin97,PethickSmith,Zapata2003,Torre2013}. We shall refer to this as a ``$J$-quench'' because $J$ is often used to denote the coupling strength between the two wells. In a simple two-mode description of a bosonic Josephson junction, i.e.\ one that assumes a single mode in each well without the quasi-continuum of low energy longitudinal modes that are present in highly elongated traps, such a quench is predicted to result in a periodic collapse and revival of the atom number distribution between the two wells \cite{Milburn1997,Tuchman2006,Chuchem2010}. Essentially the same behavior, but $\pi/2$ out of phase,  occurs in the relative phase which is the conjugate variable to number difference.  In Refs.~\onlinecite{Mumford2019,DuncanSirPRL,Goldberg2019} these revivals are shown to be examples of quantum caustics in a many-particle system. One of our main aims here is to investigate what happens to these caustics in the presence of the dispersive longitudinal modes present in the SG model, and is part of a wider program attempting to understand the role of caustics in quantum many particle dynamics \cite{DuncanSirPRL,Mumford2017,Mumford2019,Plestid2018,Goldberg2019,Kirkby2019,Kirkby2022}. 
  
  Due to the difficulty of solving the fully quantum SG model we take a semiclassical-style approach based on classical field configurations which are solutions of Eq.\ (\ref{eq:waveeqn}). Each configuration is analogous to a single geometric ray in optics and we include fluctuations by summing many configurations. The initial conditions for each field configuration are randomly sampled from a Boltzmann distribution. This approach is similar in spirit to the truncated Wigner approximation (TWA) \cite{Drummond1993,Sinatra2002,Blakie2008,QDin_phase_space,Ruostekoski2005,Javanainen2013} which includes quantum fluctuations around the classical field by summing many rays sampled from a quantum probability distribution. The TWA has previously been applied to one-dimensional Bose gases by Martin and Ruostekoski \cite{Martin2010a,Martin2010b} who studied dark solitons, and also to the connection problem of two zero temperature one-dimensional Bose gases by Dalla Torre, Demler and Polkovnikov \cite{Torre2013}, who proposed a universal scaling form for the phase dynamics after the quench. More recently, the TWA has been used  by Horv\'{a}th \textit{et al.} \cite{hungarian_paper} to  study the surprisingly sudden relaxation of the phase seen in the Vienna BEC splitting experiments \cite{Pigneur2018}. In this paper, we include both the quantum fluctuations arising from coupling two independent systems and thermal fluctuations arising from thermal phonons in the longitudinal modes and compare the time evolution of macroscopic variables (the total number difference and phase difference) in the SG system against the simpler two mode system \cite{DuncanSirPRL,Mumford2017,Mumford2019}. We find that following a quench caustics dominate the dynamics of the macroscopic variables of both systems, even in the presence of thermal fluctuations.  Due to the singular nature of caustics, and combined with their structural stability, we therefore propose that strong nongaussian fluctuations are a generic phenomenon following a quench in the SG model (and indeed, in integrable or moderately chaotic many-body systems in general).
  
  The caustics we discuss in this paper also have implications for the  question of relaxation towards equilibrium at long times in many particle systems. While chaotic (nonintegrable) and open quantum systems should thermalize (although a complete description is still the subject of active research \cite{rigol2008,dalessio2016,vznidarivc2010thermalization, purkayastha2016out,reichental2018thermalization,  tupkary2022fundamental, nathan2020universal,tupkary2023searching}), closed integrable models do not reach a conventional Gibbs state. We show here that in the SG model there is a pile-up of caustics leading to a singular shape for the long time probability distribution for the macroscopic variables that resembles the shape of a circus tent and is quite distinct from the thermal equilibrium prediction.  We find that an analytic approximation to the singular distribution based on an ergodic pendulum (assuming a microcanonical or `equal-probability' distribution) provides a good fit to the numerical data.

    The plan for the rest of this paper is as follows.    
    We start in Sec.\ \ref{sec:ham2} by deriving the SG hamiltonian from the many-body description of two coupled 1D Bose gases. In Sec.\ \ref{sec:scales} we describe the natural length and time scales and use them to write the SG hamiltonian and equations of motion in convenient dimensionless forms. Subsequently, in Sec.\ \ref{sec:IC1} we develop a method for finding the initial conditions for the SG equations of motion. We assume that prior to the quench the two Bose gases are independent and at thermal equilibrium with a bath at temperature $T$. The initial conditions are obtained by stochastically sampling the Fourier modes of a 1D quasicondensate obeying the Tomonaga-Luttinger liquid theory. With the initial conditions in hand, in Sec.\ \ref{sec:caustics_in_sine_gordon} we give the main results of this paper which are the dynamics of the macroscopic number and phase difference variables obtained by solving the equations of motion numerically.  In Sec.\ \ref{sec:importance} we consider the bigger picture and examine the universal aspects of our results including the influence of caustics on the coherence as well as the long time dynamics and the establishment of (non-thermal / non-Gaussian) equilibrium. We conclude in Sec.\ \ref{Sec:conclusion}. There are also six appendices where we give the details of the calculations as well as bench marking our numerical method.



\section{From two coupled condensates to the sine-Gordon model}\label{sec:ham2}

We begin by deriving the SG model as an effective low energy description for two coupled one-dimensional Bose gases. For the sake of clarity, we list the main simplifications employed in this work:
\begin{itemize}
    \item the treatment of a quantum many body problem by a semiclassical method (TWA). 
    \item the neglect of a weak harmonic trap along the long axis which would otherwise lead to a non-uniform longitudinal density (this can be avoided in box traps which, although rarer, can be realized \cite{Rauer2018,Navon2021})
    \item the assumption of a constant value for the tunnel coupling $J$ along the entire length of the gases
    \item the neglect of coupling to symmetric and higher transverse modes. Some more involved theoretical models do include these effects \cite{vanNieuwkerk2021,Mennemann2021}.
\end{itemize}
These simplifications are not expected to qualitatively alter the main results of this work due to the structural stability of caustics. In other words, caustics are known to be robust to perturbations in both the Hamiltonian and initial conditions. 

A theoretical description of two ultracold quasi-one dimensional gases made up of bosonic atoms of mass $m$, and held parallel to each other so that the atoms can tunnel between them at rate $J$, can be obtained from the following microscopic Hamiltonian
 \cite{Bouchoule2005,Schweigler_2017,Gritsev_2007_linear_response}
\begin{widetext}
\begin{equation}
    \begin{split}
    \hat{H}= \sum_{j=1,2} \int_{{{-L/2}}}^{{{L/2}}} dz  &  \left[-\frac{ \hbar^2}{2m} \hat{\psi}_j^{\dagger}(z) \frac{\partial^2}{\partial z^2} \hat{\psi}_j(z) +  U(z) \, \hat{\psi}_j^{\dagger}(z) \hat{\psi}_j(z)
    +\frac{g_{1D}}{2} \, \hat{\psi}_j^{\dagger}(z)\hat{\psi}_j^{\dagger}(z) \hat{\psi}_j(z)\hat{\psi}_j(z) \right] \\
    - & \int_{-L/2}^{L/2} dz \,  \hbar J\,\left[ \hat{\psi}_1^{\dagger}(z) \hat{\psi}_2(z)+\hat{\psi}_2^{\dagger}(z) \hat{\psi}_1(z)\right] \ .
        \label{eqn:Pn2}
    \end{split}
\end{equation}
\end{widetext}
The indices $j=1,2$ label the two gases and each is assumed to be tightly trapped in the $x$ and $y$ directions so that those degrees of freedom are frozen into their ground states. Only the longitudinal degree of freedom $z$ in each gas is taken to be active.  In experiments there will usually be a weak longitudinal trapping potential $U(z)$, although as mentioned above for simplicity we set it to zero and hence consider a uniform system of length $L$ with periodic boundary conditions. The quantum field operator $\hat{\psi}_j(z)$ annihilates a particle at point $z$ and together with its hermitian conjugate obeys bosonic commutation relations $[\hat{\psi}_j(z),\hat{\psi}_{j'}^{\dagger}(z')]=\delta_{jj'}\delta(z-z')$. The interaction constant $g_{1D}$ characterizes the effect of atom-atom scattering within each gas on the longitudinal degree of freedom and can be controlled both in magnitude and sign either through Feshbach or confinement-induced scattering resonances \cite{olshanii98}. An alternative physical realization of the above Hamiltonian could be a two component Bose gas in a single quasi-one dimensional trap \cite{widera2008}. In fact, quasi-zero dimensional bosonic Josephson junctions where the atoms are held in a single tight trap and two atomic spin states are used for the two states have been studied experimentally \cite{zibold10}. 

A weakly interacting three-dimensional Bose gas at ultracold temperatures will undergo Bose-Einstein condensation and can be described to high accuracy by a classical field approximation (Gross-Pitaevskii theory \cite{dalfovo99}). In a quasi-one dimensional geometry quantum fluctuations can still be small if the density is not too low, and under these circumstances the gas can be treated as a quasicondensate where the  quantum field operators are replaced by classical fields  \cite{quantum_regimes_paper,sinatra01,mora03}
\begin{equation}
  \hat{\psi}_{j}(z) \rightarrow  \psi_{j}(z)=  \exp[i \phi_{j}(z)] \, \sqrt{n_{1D}+\rho_{j}(z)} \ .
        \label{eqn:Pn3}
\end{equation}
Here $n_{1D}=N/L$ is the background density where $N$ is the number of atoms in each gas (for simplicity we assume an equal number of atoms $N$ in each gas; the structural stability of caustics means that they are in fact stable to small differences in $n_{1D}$ between the two gases as we will see in Section \ref{sec:structuralstability}). $\rho_{j}(z)$ and $\phi_{j}(z)$ are the atom number density and phase fluctuations at each point $z$, respectively. These are canonically conjugate variables and can even be quantized in a semiclassical regime such that they obey the commutation relations $[\hat{\rho}_j(z),\hat{\phi}_{j^{'}}(z')] \approx i \delta_{jj'}\delta(z-z')$ in a coarse grained sense \cite{mora03}. However, in the present paper $\rho_{j}(z)$ and $\phi_{j}(z)$ will be purely classical fields subject only to thermal fluctuations. 

We can further decompose the fields into their symmetric and antisymmetric components
\begin{eqnarray}
     \rho_s(z) = \frac{\rho_{1}(z)+\rho_{2}(z)}{2}, \quad \rho_a(z) = \frac{\rho_{1}(z)-\rho_{2}(z)}{2} \nonumber \\
     \phi_s(z) = \phi_{1}(z)+\phi_{2} (z), \quad \phi_a(z) = \phi_{1}(z)-\phi_{2}(z) \ . \label{eq:symmetric_antisymmetric}
\end{eqnarray}
In terms of these variables the Hamiltonian in Eq.\ (\ref{eqn:Pn2}) separates into three parts: a part that depends only on the antisymmetric variables, a part that depends only on the symmetric variables, and a single term  $-\hbar J\int dz \, \rho_{s}(z) \cos \phi_{a}(z)$ that couples the two \cite{Bouchoule2005,Gritsev_2007_linear_response,Schmiedmayer,Gring_and_Schmeidmayer_relaxation_paper,Schweigler_2017,Schmeidmayer_double_light_cone_paper}. 
However, provided the background density $n_{1D}$ of the two gases is the same and the system is at or close to thermal equilibrium so that the  density and phase fluctuations are uncorrelated (which is the initial condition we assume in this paper), the coupling term vanishes so that the symmetric and antisymmetric parts of the Hamiltonian decouple  \cite{Gritsev_2007_linear_response,Schweigler_2017,Schmeidmayer_double_light_cone_paper}. We can therefore restrict attention to either the symmetric or the antisymmetric sectors. In this paper, we focus entirely on the antisymmetric sector because the $J$-quench we propose to launch the dynamics only couples to the antisymmetric variables and fortunately these are also the variables most easily measured in the matter wave interference measurement method favored in experiments \cite{Hofferberth2008,Betz2011,Gring_and_Schmeidmayer_relaxation_paper,Local_emergence_and_correlation_paper,Schweigler_2017,Schmeidmayer_double_light_cone_paper,Rauer2018,Pigneur2018}.

The Hamiltonian describing the antisymmetric variables that one obtains from Eq.\ (\ref{eqn:Pn2}) is (see Appendix \ref{app:sine_gordon_derivation} for details)
\begin{equation}
\begin{split}
    H_{\mathrm{SG}+}  = & \int_{-L/2}^{L/2}  dz \bigg[g_{1D} \, \rho_{a}^{2}(z) + \frac{\hbar^2 n_{1D}}{4m}\left(\frac{\partial \phi_a}{\partial z}\right)^2  \\ & +  \frac{\hbar^2}{4m n_{1D}}\left(\frac{\partial \rho_a}{\partial z}\right)^2 
    -  2 \hbar J n_{1D} \cos \phi_a(z) \bigg] \ .
    \label{eq:fullassym}
\end{split}
\end{equation}
We refer to this as the ``sine-Gordon plus'' (SG+) model because it includes an extra term (the third term) in comparison to the standard SG Hamiltonian 
\begin{equation}
\begin{split}
    H_{\mathrm{SG}}= \int_{-L/2}^{L/2}  dz \,  \bigg[ & g_{1D} \,  \rho_a(z)^2 + \frac{\hbar^2 n_{1D}}{4m} \left(\frac{\partial \phi_a}{\partial z}\right)^2  \\
     & -   2 \hbar J n_{1D} \cos \phi_a(z) \bigg] \ .
\label{eq:Pn1}
\end{split}
\end{equation}
The third term in the SG+ model involves the gradient of density fluctuations and has the effect of suppressing density fluctuations at small length scales which are otherwise free to proliferate.  Taking the ratio of the density fluctuations (the first term) to the gradient of the density fluctuations (the third term) we see that the intrinsic length scale associated with density fluctuations is the healing length 
\begin{equation}
\xi_h=\frac{\hbar}{\sqrt{m g_{1D}n_{1D}}  
} 
\label{eq:healinglength}
\end{equation} 
which represents the minimum length at which our theory in terms of classical density and phase fields is valid \cite{ Gritsev_2007_linear_response,hungarian_paper}. Even if short wavelength modes are not excited initially, the nonlinearity of the SG model couples the different modes and over time they can become excited unless some kind of regularization, such as occurs naturally through the third term in the SG+ model, is applied.

As shown in chapter 3 of the monograph in reference \cite{Giamarchi_book}, renormalization group treatments show in fact that the third term in Eq.\ (\ref{eq:fullassym}) is formally irrelevant, but in order to prevent small scale density fluctuations in numerical calculations previous authors have used a lattice where the spacing is chosen to be greater than $\xi_{h}$ \cite{hungarian_paper}. In our work we implement both the lattice regularization procedure and retain the third term in Eq.\ (\ref{eq:fullassym}) so that the cut off is applied smoothly.

 The nonlinear term in the SG and SG+ Hamiltonians is the cosine term which originates from tunneling between the two wells and occurs in all Josephson junction type problems. As noted above this is also the only term that is directly modified by the $J$-quench as  there is no term depending on $J$ in the symmetric sector \cite{Bouchoule2005,Gritsev_2007_linear_response}. From the structure of this term it can be seen that it appears as a potential well for phase configurations $\phi(z,t)$. In fact, as we shall see in Section \ref{sec:caustics_in_sine_gordon}, it behaves like an (imperfect) lens that focuses such phase `rays' excited by the quench to form caustics.
  For the sake of brevity, and when we deem no confusion can arise, we will  omit the `$a$' subscript on antisymmetric variables since we will not be dealing with symmetric degrees of freedom.

The fact that the two fields $\phi(z)$ and $\rho(z)$ form a conjugate pair means that their equations of motion are given by Hamilton's equations
\begin{equation}
\begin{split}
	\dot{\phi} &= \frac{1}{\hbar}\frac{\delta \mathcal{H}}{\delta  \rho(z)}\\
	 \dot{\rho} &= - \frac{1}{\hbar}\frac{\delta \mathcal{H}}{\delta \phi(z)} 
\label{eq:variational_eqn1}
\end{split}
\end{equation}
where $\mathcal{H}$ is the Hamiltonian density defined via 
\begin{equation}
H=\int_{-L/2}^{L/2} {\mathcal{H} \,dz}.
\end{equation}
Applying these equations to the SG+ Hamiltonian given in Eq.\ (\ref{eq:fullassym}) we find the following of equations of motion 
\begin{equation}
\begin{split}
	\frac{d\phi(z,t)}{dt} &=  2 \ \frac{g_{1D}}{\hbar} \rho(z,t)+2 \ \frac{\hbar}{4m n_{1D}}\frac{\partial^2 \rho(z,t)}{\partial z^2}\\
	\frac{d\rho(z,t)}{dt} &=  2 \ \frac{\hbar n_{1D}}{4m} \frac{\partial^2 \phi(z,t)}{\partial z^2}-2 J n_{1D} \sin[\phi(z,t)] \ .
\label{eq:EOM1}
\end{split}
\end{equation}
These are the key equations we use to solve for the dynamics of the field configurations. They have the form of Josephson's equations \cite{smerzi97} augmented by second order spatial derivatives  $\partial^2 \phi/\partial z^2$ and $\partial^2 \rho/\partial z^2$ which account for phase and density fluctuations along the longitudinal direction. Combined with the sine term, they will cause wavepackets to disperse along $z$.    In the absence of these terms we have exactly the equations of motion for a  pendulum where $\phi$ is the angular displacement from equilibrium and $\rho$ plays the role of angular momentum. The dependence on $z$ suggests an interpretation in terms of a continuous chain of many pendula each coupled to its neighbors by the spatial derivative terms and is reminiscent of the Fermi-Pasta-Ulam-Tsingou problem \cite{Bouchoule2005,Fermi1955}.

In this paper the coupled equations of motion given in Eq.\ (\ref{eq:EOM1}) will be solved numerically for a system of length $L$. To perform the numerical computations we discretize the system on a spatial grid with $N_L+1$ points which makes the grid spacing $a  = L/N_L$.
The positions of the grid points are given by $z=ra$ where $r$ is an integer in the range
\begin{equation} 
r=-\frac{N_L}{2},\ldots,\frac{N_L}{2}
\label{eq:numericalgridpoints}
\end{equation}
and $N_L$ is chosen to be an even integer.
In light of the discussion above concerning the role of the healing length as a physical cut off, we follow reference \cite{hungarian_paper} and perform our numerics on lattices with grid size $a$ greater than $\xi_h$. 
This implies 
\begin{equation}
N_{L}^2 < \frac{m g_{1D} n_{1D} L^2}{\hbar^2}. 
\label{eq:NLcondition}
\end{equation}
We fulfil the condition given in Eq.~\eqref{eq:NLcondition} in all our numerics. In Appendix \ref{sec:benchmarking} we check and verify the convergence of our numerical solution of  Eq.~\eqref{eq:EOM1} as a function of the grid size.

\section{Natural scales}
\label{sec:scales}

Let us express the SG/SG+ Hamiltonians and equations of motion in terms of the natural scales for a one-dimensional quantum fluid. For a length scale we chose the healing length $\xi_{h}$ given in Eq.\ (\ref{eq:healinglength}). The ratio of the healing length to the mean interparticle spacing $1/n_{1D}$ gives rise to the dimensionless parameter
\begin{equation}
 K = \sqrt{\frac{n_{1D} (\hbar \pi)^2}{ 4 g_{1D} m} } 
\label{eq:cK}
\end{equation}
which measures how strongly interacting the system is. When $K \gg 1$ the healing length is much greater than the interparticle spacing and the system is in the weakly interacting (quasicondensate) regime. In this limit we can identify $K$ with the Luttinger parameter from Tomonaga-Luttinger (TL) theory
that provides the universal low energy effective description for one-dimensional quantum fluids \cite{Haldane1981} (low energy limit of the Lieb–Liniger theory, for example \cite{Jiang2015}) and that has been applied to quasi-one dimensional ultracold atomic gases by a number of authors, e.g.\ \cite{Gritsev_2007_linear_response,Torre2013,hungarian_paper,Bouchoule12,Giamarchi_book,Didier2009,Imambekov2012}. The  relationship between $K$ and microscopic quantities such as $m$ and $g_{1D}$ is not known analytically in the general case, but in the weakly interacting limit it reduces to Eq.\ (\ref{eq:cK}). The same is true of the speed of sound which here is given by 
\begin{equation}
c = \sqrt{\frac{g_{1D} n_{1D}}{ m }} .
\label{eq:speedofsound}
\end{equation}
This can be used to define a characteristic energy, namely that associated with phonons (quanta of sound)
\begin{equation}
E=\hbar \omega=\frac{\hbar c}{\xi_{h}} 
\end{equation}
where we have set the natural frequency  $\omega$ to be the ratio of the speed of sound to the healing length.

 We therefore transform to the following dimensionless variables
\begin{equation}
    \begin{split}
     z &\xrightarrow{} \Tilde{z}=\frac{z}{\xi_{h}} \quad , \quad t \xrightarrow{} \tilde{t}= \frac{c}{\xi_{h}} t \\
     \rho &\xrightarrow{}  \Tilde{\rho}=\rho \; \xi_{h} \quad , \quad
     \phi \xrightarrow{} \Tilde{\phi}=\phi
     \label{eq:dimensionless_form}
    \end{split}
\end{equation}
and defining $\tilde{H}_{\mathrm{SG}}=H_{\mathrm{SG}}/E$ and likewise for $\tilde{H}_{\mathrm{SG}+}$  we obtain the two Hamiltonians in dimensionless form
\begin{equation}
    \tilde{H}_{\mathrm{SG}}=\int_{-L/2}^{L/2}  d \tilde{z} \,  \bigg[ \Gamma \, \tilde{\rho}^2 + \epsilon \left(\frac{\partial \tilde{\phi}}{\partial \tilde{z}}\right)^2 
    -2 \mathcal{J} \cos \tilde{\phi} \bigg]
    \label{eq:HSG_dimensionless}
\end{equation}
and
\begin{eqnarray}
    \tilde{H}_{\mathrm{SG}+} &=& \int_{-L/2}^{L/2}  d \tilde{z} \,  \bigg[ \Gamma \, \tilde{\rho}^2 + \epsilon \left(\frac{\partial \tilde{\phi}}{\partial \tilde{z}}\right)^2 + \frac{\Gamma}{4}\left( \frac{\partial \tilde{\rho}}{\partial \tilde{z}} \right)^2 \nonumber \\
    &-& 2 \mathcal{J} \cos \tilde{\phi} \bigg]
    \label{eq:HSGplus_dimensionless}
\end{eqnarray}
where the coefficients are given by
\begin{equation}
        \Gamma  = \frac{\pi}{2K} \ , \
        \epsilon  = \frac{K}{2\pi} \ , \
        \mathcal{J}  = \frac{K}{2 \pi}\frac{\xi_{h}^2}{\xi_{s}^2} \ .
        \label{eq:coefficientlist}
\end{equation}
In the last term we have introduced the spin healing length
\begin{equation}
    \xi_{s}=\sqrt{\frac{\hbar}{4mJ}} 
\end{equation}
which provides a measure for the distance over which coherence between the two gases is restored due to the tunnel coupling $J$ \cite{Schmiedmayer}. At finite temperatures another useful length scale is the thermal phase coherence length 
\begin{equation}
\lambda_{T}=\frac{2\hbar^2 n_{1D}}{m k_{B}T}.
\end{equation}

The dimensionless form of the equations of motion can now be given.  For the SG model we find 
\begin{equation}
\begin{split}
    \frac{d \tilde{\phi}}{d \tilde{t}} & =2 \Gamma \tilde{\rho} \\
    \frac{d \tilde{\rho}}{d \tilde{t}} & = 2 \epsilon \frac{\partial^2 \tilde{\phi}}{\partial \tilde{z}^2} - 2 \mathcal{J} \sin \tilde{\phi} 
\end{split}
\label{eq:eomdimensionless}
\end{equation}
and for the SG+ model we obtain
\begin{equation}
\begin{split}
    \frac{d \tilde{\phi}}{d \tilde{t}} & =2 \Gamma \tilde{\rho} - \frac{\Gamma}{2} \frac{\partial^2 \tilde{\rho}}{\partial \tilde{z}^2}  \\
    \frac{d \tilde{\rho}}{d \tilde{t}} & = 2 \epsilon \frac{\partial^2 \tilde{\phi}}{\partial \tilde{z}^2} - 2 \mathcal{J} \sin \tilde{\phi} \ .
\end{split}
\label{eq:eomSGplusdimensionless}
\end{equation}

\section{Initial Conditions}\label{sec:IC1}
The dynamics we seek to study in this paper start from
a $J$-quench where two independent one-dimensional gases
at thermal equilibrium are suddenly coupled. In order
to obtain the initial density and phase fluctuations of
these gases we use the TL model.

\subsection{Tomonaga-Luttinger (TL) liquid}
\label{subsec:TL}

In our notation the TL Hamiltonian reads
\begin{equation}
\begin{split}
    H_{\mathrm{TL}}&= \int_{-L/2}^{L/2}   { dz \left[g_{1D} \rho_{j}(z)^2 + \frac{\hbar^2 n_{1D}}{4m}\left(\frac{\partial \phi_{j}}{\partial z}\right)^2 \right] } 
\label{eq:H_TL}
\end{split}
\end{equation}
where $j$ labels either of the two gases. We henceforth, omit this label for the sake of brevity with the understanding that in this section the density and phase fields refer to just one of the two gases. Eq.~\eqref{eq:H_TL} has the same mathematical structure as the SG model but without the tunnelling term. If we include the gradient of density fluctuations we can define
\begin{equation}
\begin{split}
    H_{\mathrm{TL}+}= \int_{-L/2}^{L/2}    dz \bigg[g_{1D} \rho_{}(z)^2  & + \frac{\hbar^2 n_{1D}}{4m}\left(\frac{\partial \phi_{}}{\partial z}\right)^2 \\  & + \frac{\hbar^2}{4m n_{1D}}\left(\frac{\partial \rho_{}}{\partial z}\right)^2 \bigg] \ .
\label{eq:H_TLplus}
\end{split}
\end{equation}
The TL model is quadratic and hence its thermal fluctuations can be treated exactly. To this end it is useful to work in Fourier space and we apply discrete Fourier transforms defined on the numerical grid with $N_{L}$ points as discussed at the end of Section \ref{sec:ham2}. The phase field $\phi$ and its Fourier transform $\varphi$ are related by 
\begin{equation}
\begin{split}
 &\phi_r=\frac{1}{\sqrt{N_{L}+1}}\sum_{k=-N_{L}/2}^{N_L/2}\varphi_{k} \ \exp \left[i \frac{2\pi k r}{N_L+1} \right] \\
  &\varphi_k=\frac{1}{\sqrt{N_{L}+1}}\sum_{r=-N_{L}/2}^{N_L/2}\phi_{r} \ \exp \left[-i \frac{2\pi k r}{N_L+1} \right] \ .
  \label{fourier_modes_convenient_formulation}
\end{split}
\end{equation}
The discrete data $\{\phi_{r}\}=\{ \phi_{-N_{L}/2},\ldots, \phi_{0}, \ldots , \phi_{N_{L}/2} \}$ and its transform are located symmetrically about $r=0$ and $k=0$, respectively. Since the value $\phi_{r}$ of the field at each coordinate space grid point is a real number the  condition \ $\varphi_{-k}=\varphi_{k}^{\ast}$ must hold. Similarly the density fluctuation field $\rho$ and its Fourier transform $\varrho$ are related by
\begin{equation}
\begin{split}
 &\rho_r=\frac{1}{\sqrt{N_{L}+1}}\sum_{k=-N_{L}/2}^{N_L/2}\varrho_{k} \ \exp \left[i \frac{2\pi k r}{N_L+1} \right]\\
  &\varrho_k=\frac{1}{\sqrt{N_{L}+1}}\sum_{r=-N_{L}/2}^{N_L/2}\rho_{r} \ \exp \left[-i \frac{2\pi k r}{N_L+1} \right] 
  \label{eq:Fourier_summation}
\end{split}
\end{equation}
where again the reality of the field in coordinate space  requires that $\varrho_{-k}=\varrho_{k}^{\ast}$.
Inserting these transformations in Eq.\ (\ref{eq:H_TLplus}) we obtain (see Appendix \ref{appendix:hamiltonian initial condition} for details)
\begin{equation}
\begin{split}
    H_{\mathrm{TL}+}&= a  \, g_{1D} \,  \sum_{k=-N_L/2}^{N_L/2} \, |\varrho_k|^2    \\
    &+ a \ \hbar \ n_{1D} \sum_{k=-N_L/2}^{N_L/2} \frac{\hbar \pi^2 k^2 }{m L^2  } \ |\varphi_k|^2  \\
    & + a \ \frac{\hbar^2}{4m n_{1D}}  \sum_{k=-N_L/2}^{N_L/2} \frac{ 4 \pi^2 k^2 }{L^2  } \ |\varrho_k|^2 \ .
\end{split}
\label{Sine-Gordon hamiltonian fourier form plus}
\end{equation}
Before proceeding with further analysis of Eq.~\eqref{Sine-Gordon hamiltonian fourier form plus}, it is worth noting that it can be recast in a standard Luttinger liquid form
\begin{equation}
\begin{split}
    H_{\mathrm{TL}+} = \frac{a c \hbar}{2}   \sum_{k=-N_L/2}^{N_L/2} \, \bigg[ \frac{K}{\pi} & \frac{4\pi^2 k^2}{L^2}  |\varphi_k|^2  + \frac{\pi}{K} |\varrho_k|^2  \\
    & +  \frac{K}{\pi} \frac{4 \pi^2 k^2}{N^2} |\varrho_k|^2 
    \bigg]   
\end{split}
\label{cK_Sine-Gordon hamiltonian fourier form}
\end{equation}
where the strength of the terms depends either on $K$ or $1/K$.

Applying the transformations given in Eq.\ (\ref{eq:dimensionless_form}), the Fourier space variables can be written in dimensionless form as
\begin{equation}
    \begin{split}
     \varrho_k \xrightarrow{}  \Tilde{\varrho}_k=\xi_{h} \varrho_k \quad , \quad 
     \varphi_k \xrightarrow{} \Tilde{\varphi}_k=\varphi_k \ 
     \label{eq:dimensionless_form_Fourier_modes}
    \end{split}
\end{equation}
 and the TL+ Hamiltonian given in Eq.~\eqref{Sine-Gordon hamiltonian fourier form plus} scaled by the energy $E=\hbar c /\xi_{h}$  is given by
 
\begin{equation}
\begin{split}
    \tilde{H}_{\mathrm{TL}+}=   \frac{\tilde{L}}{N_L}\sum_{k=-N_L/2}^{N_L/2} \, \bigg[  & \frac{\epsilon \,  4 \pi^2 k^2}{\tilde{L}^2}  |\tilde{\varphi}_k|^2  +   \Gamma |\tilde{\varrho}_k|^2  \\
    & +   \frac{\Gamma \,  \pi^2 k^2}{\tilde{L}^2}  \vert \tilde{\varrho}_k \vert^2 \bigg]  
\end{split}
\label{tilde_cK_Sine-Gordon hamiltonian fourier form plus}
\end{equation}
where $\tilde{L}=L/\xi_{h}$ is the ratio of the system size to the healing length. Comparison with the spatial version of $H_{\mathrm{TL+}}$ given in Eq.\ (\ref{eq:H_TLplus}) shows where this factor comes from: as the size is increased the range of the integration increases linearly and this is accounted for by $\tilde{L}$ in the Fourier transformed version. Note that all parameters and variables in Eq.~\eqref{tilde_cK_Sine-Gordon hamiltonian fourier form plus} are dimensionless.

\subsection{Thermal equilibrium}\label{subsec:IC2}

To find the initial conditions on the fields  $\rho_{j}(z)$ and $\phi_{j}(z)$ we assume that each gas is at thermal equilibrium such that the excitation (phonon) modes of the TL+ Hamiltonian are populated with a probability given by the Boltzmann distribution. The range of temperatures we simulate is listed in Table \ref{parameter_table} along with the values of all the other key parameters,  and is chosen so as to correspond to realistic experimental conditions. For our theoretical treatment to be valid the temperature must be below that where the quasicondensate state occurs. This temperature is of the order of $T_{\mathrm{qc}} \sim \sqrt{g_{1D}\hbar^2 n_{1D}^3/m}/k_{B}$ \cite{Kheruntsyan2005,Bouchoule2007,Wang2013}. Substituting in the values of the parameters found Table \ref{parameter_table} we find $T_{\mathrm{qc}} \sim 1 \mu$K which is several orders of magnitude higher than the temperatures we choose in this paper.

 In the canonical ensemble of statistical mechanics the probability that a system at thermal equilibrium has the phase space configuration $s = {q_1, p_1, q_2, p_2...q_N , p_N }$ is proportional to the Boltzmann weight $\exp[- \beta H(s)]$, where $\beta=1/k_{B}T$ and $H=\sum_i{p_i^2/2m+V(q_i)}$. The Hamiltonian in Eq.\ (\ref{tilde_cK_Sine-Gordon hamiltonian fourier form plus}) is quadratic and hence the Boltzmann weight becomes that of a series of independent harmonic oscillators
\begin{equation}
\begin{split}
    e^{-\tilde{\beta} \Tilde{H}_{\mathrm{TL+}}}=\prod_{k} e^{-  P_{k}^{2}/2 \sigma^{2}_{\rho +}} \ e^{- Q_{k} ^{2}/2 \sigma_{\phi +}^2(k) }
\label{eq:Pn12}
\end{split}
\end{equation}
where  {$\tilde{\beta}= (\hbar c/\xi_{h}) /k_{B}T $ is the appropriately scaled temperature parameter} 
and we have introduced the real variables $Q_{k}$ and $P_{k}$ which are related to the old variables by 
{\begin{equation}
\tilde{\varphi}_k=Q_{k}e^{i \alpha_k}, \, \quad\, \tilde{\varrho}_k= P_{k}e^{i \beta_k}. 
\label{eq:QkPk}
\end{equation}}
The phases $\alpha_{k}$ and $\beta_{k}$ allow for the fact that $\tilde{\varphi}_k$ and $\tilde{\varrho}_k$ can be complex numbers. The variances in Eq.\ (\ref{eq:Pn12}) are given by
{\begin{eqnarray}
\sigma^2_{\rho+}(k) & = & \frac{N_{L}}{ 2 \tilde{\beta} }\frac{1}{\Gamma \tilde{L}(1+\pi^2 k^2/\tilde{L}^2)} \label{eq:sigmarhoplus} \\ \sigma^2_{\phi+}(k)  & = & \frac{N_{L}}{2\Tilde{\beta}}  \frac{\tilde{L}}{4\pi^2 k^2  \epsilon} \ . \label{eq:sigmaphiplus}
\end{eqnarray}
}
The partition function can now be written down as
\begin{equation}
\begin{split}
 Z= & \prod_{k} \int_{-\infty}^{\infty}   e^{-\tilde{\beta}  \Tilde{H}_{\mathrm{TL}+}} \  dP_{k} dQ_{k}\\
  & =  \prod_{k} \left(\sigma_{\rho +}\sqrt{2\pi}\right)\left(\sigma_{\phi +}(k)\sqrt{2\pi}\right) 
 \label{eqn:partition_function}
 \end{split}
\end{equation}
and hence the probability $\mathcal{P}$ of a particular configuration $(Q_1,Q_2,....,P_1,P_2,....)$ is
\begin{equation}
    \mathcal{P}=\prod_{k}\left(\frac{e^{-P_{k}^{2}/2 \sigma^{2}_{\rho +}}}{ \sigma_{\rho +}\sqrt{2\pi}} \right) \left(\frac{e^{-Q_{k}^{2}/2 \sigma_{\phi +}^2(k)}}{\sigma_{\phi +}(k)\sqrt{2\pi}}\right) \ .
\label{eq:probability_distribution}
\end{equation}
This is seen to be the total probability distribution for independent random variables $P_{k}$ and $Q_{k}$ drawn from normal distributions.
{Thus, the absolute values of the Fourier coefficients $\tilde{\varrho}_k$ and $\tilde{\varphi}_
k$ are normally distributed random variables with zero mean and variances given by Eqns.\ (\ref{eq:sigmarhoplus}) and (\ref{eq:sigmaphiplus}). We sample these numerically from normal distributions to generate the initial system configuration. {The phases $\alpha_{k}$ and $\beta_{k}$} given in Eq.\ (\ref{eq:QkPk}) do not appear in the Boltzmann weight and are  chosen randomly from the range $[0,2 \pi)$. In fact, for both the phases and the amplitudes we only need to choose the values for terms with $k \ge 0$ because the reality conditions imply that we can put
\begin{eqnarray}
   Q_k=Q_{-k} \ , \quad P_k=P_{-k}, \nonumber\\
    \alpha_k=-\alpha_{-k} \ , \quad \beta_k=-\beta_{-k} \ .
    \label{eqn:sampling_modes}
\end{eqnarray}
}


So far we have only considered the initial state of a single gas. By subtracting the results for two gases we can obtain the initial values of the antisymmetric variables $\rho_{a}(z)$ and $\phi_{a}(z)$ defined in Eq.\ (\ref{eq:symmetric_antisymmetric}). Actually, due to the fact that the SG+ Hamiltonian with $J=0$ and expressed in terms of antisymmetric variables as given in Eq.\ (\ref{eq:fullassym})  formally has the same structure as the TL+ Hamiltonian given in Eq.\ (\ref{eq:H_TLplus}),  sampling initial data for two gases is unnecessary and one can obtain $\rho_{a}(z)$ and $\phi_{a}(z)$ directly by sampling them as though they were from one gas described by the TL+ Hamiltonian. However, in doing so, consideration needs to be given to the average value of relative phase $\phi_{a}(z)$ because both the SG+ and TL+ Hamiltonians only contain the spatial derivative of the phase but not the phase itself. Its average value is therefore not determined by energy considerations and is left to float freely. This is also apparent in the Fourier transformed version of the TL Hamiltonian given in Eq.\ (\ref{tilde_cK_Sine-Gordon hamiltonian fourier form plus}) where the $k=0$ term involving $\tilde{\varphi}_{0}$ is absent due to the vanishing of its coefficient which is proportional to $k^2$. To take into account the random phase difference between the two gases one can chose $\tilde{\varphi}_{0}$ to be a random number in the range $[-\pi\ldots \pi)$ but multiplied by a factor of $\sqrt{N_{L}+1}$ in order to respect the normalization in  Eq.\ (\ref{fourier_modes_convenient_formulation}). This gives  values of the average value of $\phi_{a}(z)$ in the desired range $-\pi$ and $+\pi$.

The random value of the initial phase difference is actually a key feature of the $J$-quench. It populates the cosine potential landscape in the Hamiltonian with uniform probability. As the trajectories roll back and forth in this potential they form caustics. In effect, the cosine potential acts as an imperfect lens that focuses an initially flat `wavefront' over time.

\begin{center}
\begin{table}[t]
\begin{tabular}{||c c c||} 
 \hline
 Symbol & Parameter & Value  \\ [0.5ex] 
 \hline\hline
 $\omega_{\perp}$ & trapping frequency & $2\pi \times 3$ kHz \\
 \hline
 $m$ & mass of Rb atom & $ 1.41\times10^{-25} $ kg \\
 \hline
 $a_{s}$ & scattering length & $ 98\times0.52$ \AA  \\
 \hline
 $N$ & number of atoms & 1200  \\
 \hline
 $L$ & system length & 18 $\mu$m  \\
 \hline
 $n_{1D}$ & average density & $6.7 \times 10^7 \mathrm{m}^{-1}$\\
 \hline
 $g_{1D}$ & 2 $\hbar a_{\mathrm{scat}} \omega_{\perp}$ &  $2\times 10^{-38}$ Jm \\ 
 \hline
 $K$ & Luttinger parameter & 25 \\
 \hline
 $T$ & temperature & 2 - 20 nK \\
 \hline
 $J$ & $J$-quench & 0\ -\ 30 Hz \\
 \hline
 $N_{L}$ &number of grid points & 50  \\ 
 \hline
  $c$ & speed of sound & $3 \times 10^{-3}$ m s$^{-1}$ \\ 
  \hline
 a & grid spacing & $0.36 \ \mu$m. \\
 \hline
 $\xi_h$ & healing length & $0.24 \ \mu$m \\
 \hline
$\lambda_T$ & phase coherence length & $ 38 - 380 \ \mu$m\\
\hline
$\xi_s$ & spin healing length & $2.5 \ \mu$m\\
\hline

\end{tabular}
 \caption{Table containing important parameters and their values. The parameters are chosen to be experimentally feasible and correspond roughly to those reported in references \cite{Gring_and_Schmeidmayer_relaxation_paper,Schmeidmayer_double_light_cone_paper,Schweigler_2017,Local_emergence_and_correlation_paper,Rauer2018,Pigneur2018}.}
 \label{parameter_table}
 \end{table}
 
\end{center}

\begin{figure*}[t]
\centering
\begin{tabular}{ccc}
   \includegraphics[scale = 0.37]{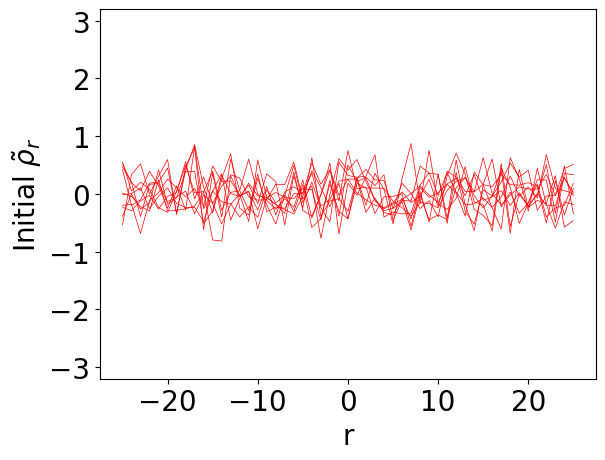} &     
   \includegraphics[scale = 0.37]{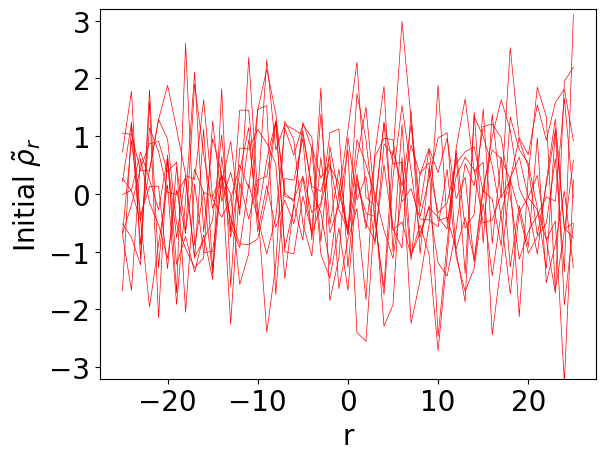} &
   \includegraphics[scale = 0.37]{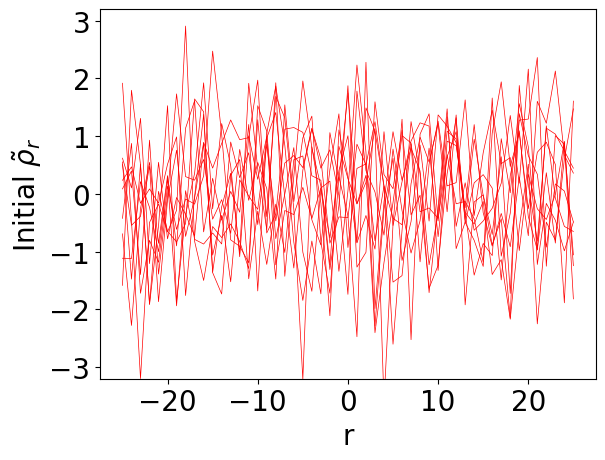} \\
  \includegraphics[scale = 0.37]{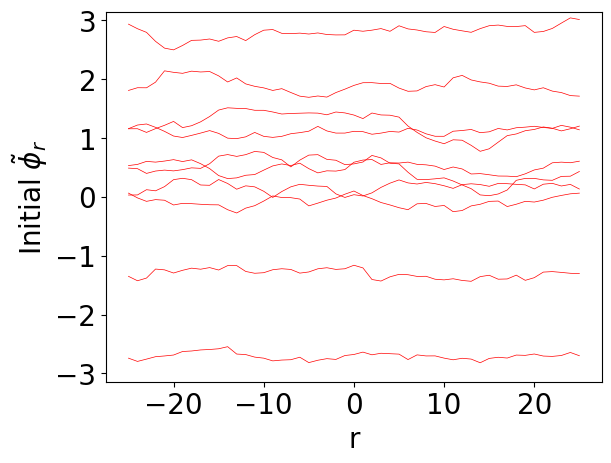} &
    \includegraphics[scale = 0.37]{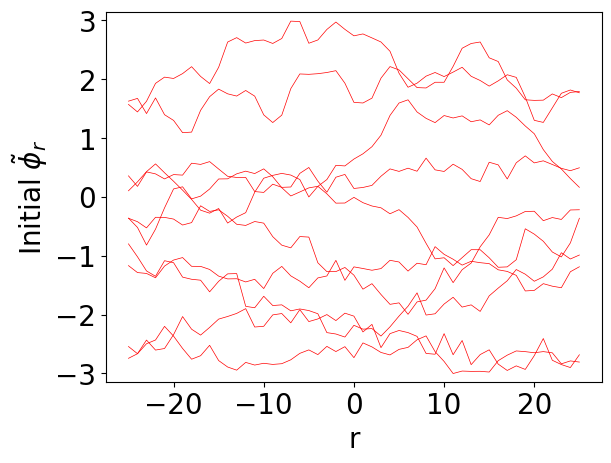} &
 \includegraphics[scale = 0.37]{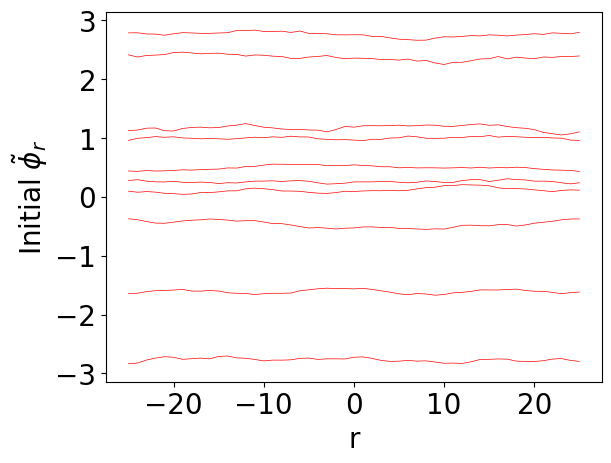}
 \end{tabular}
\caption{\label{fig:my-label5} Examples of initial spatial profiles of the number difference $\tilde{\rho}$ (top row) and phase difference $\tilde{\phi}$ (bottom row) plotted as a function of the computational lattice points $r$. Each profile is obtained by randomly sampling a thermal distribution using the method described in Section \ref{subsec:IC2}, and each panel includes ten different profiles. The parameter values common to all panels include the number of computational lattice points $N_L=50$, grid spacing $a=0.36\mu m$, and healing length $\xi_h= 0.24\,\mu$m (the remaining parameters are listed in  Table \ref{parameter_table}). The difference between the columns is as follows. 
The left column has the Luttinger parameter $K=25$, and temperature $T=2$ nK giving a phase coherence length of $\lambda_{T}=380\,\mu$m. In the middle column $K=25$, but the temperature is increased to $20\,$nK, giving $\lambda_{T}=38\,\mu$m. In the right column, the value of $K$ is artificially increased (without changing any other parameters) to $K=250$ and $T=2$ nK. Increases in temperature excite stronger fluctuations in the profiles as expected. Increases in the Luttinger parameter have opposite effects on $\tilde{\rho}$ and  $\tilde{\phi}$.  The maximum value and jaggedness of $\Tilde{\rho}$ is increased whereas the jaggedness of $\tilde{\phi}$ is reduced. An explanation of this behavior is given in the main text.}
\end{figure*}      



\subsection{Choice of parameters}\label{sec:wir}

There are three constraints which must be satisfied in order to have a quasi-one dimensional condensate \cite{Schmiedmayer}. To ensure minimal scattering into the transverse modes we need the interaction to be sufficiently weak in comparison to the transverse trapping potential which implies $\mu= g_{1D}n_{1D} \ll \hbar \omega_{\perp}$ where $\mu$ is the chemical potential and $\omega_{\perp}$ is the transverse trapping frequency. The temperature must also be low enough that transverse modes are not thermally excited leading to the inequality $k_{B} T \ll \hbar \omega_{\perp}$ (this turns out to be more stringent than the quasicondensate temperature $T_{\mathrm{qc}}$ mentioned in Section \ref{subsec:IC2}). Finally, even if we are below $T_{\mathrm{qc}}$, a quasicondensate also requires weak interactions in comparison to the zero-point kinetic energy associated with the density of the particles. This implies $n_{1D} g_{1D} \ll \hbar^2 n_{1D}^{2}/m$ which means the Luttinger parameter should obey $K \gg 1$. All the parameter values we use satisfy these three inequalities.

In quasi-one dimensional gases the interatomic interaction parameter $g_{1D}$ is related to the scattering length $a_{s}$ and transverse trapping frequency as  $g_{1D}=2\hbar a_s \omega_{\perp}$. For $^{87}$Rb atoms we have  $a_{s} \approx 98 \times 0.52\, $\AA    \cite{s_wave_scattering_paper} and we will assume
$\omega_{\perp}=2 \, \pi \times$3 kHz \cite{Pigneur2018}.  The full list of parameters used in our simulations is given in Table \ref{parameter_table} and roughly  corresponds to those used in the experiments by the Vienna group
\cite{Gring_and_Schmeidmayer_relaxation_paper,Schmeidmayer_double_light_cone_paper,Schweigler_2017,Local_emergence_and_correlation_paper,Rauer2018,Pigneur2018}.

For our numerical simulations we choose a grid size that slightly exceeds the healing length because, as explained above, this cuts off unphysical density fluctuations \cite{Gritsev_2007_linear_response,hungarian_paper}.  This condition is given in Eq.~(\ref{eq:NLcondition}) but can be expressed succinctly in terms of $\Gamma$ as $N_{L}^2 < \Gamma N^2$. The magnitudes of $\tilde{\rho}$ and $\tilde{\phi}$ also need to be considered. The phase difference can take the full range $+\pi$ to $-\pi$, but the number difference is limited by the condition that the total number difference (integrated over the entire system) cannot exceed the total number of particles. In fact, due to the random nature of sampled thermal fluctuations, the integral of $\tilde{\rho}$ is always approximately zero. However, the validity of the SG/SG+ model requires that local density fluctuations be small in comparison to the background density $n_{1D}$,  see Appendix \ref{app:sine_gordon_derivation}. Translated into the scaled variables this means that at any point $\Tilde{\rho}(\tilde{z}) \ll n_{1D}\xi_h$. In practice we choose  $\Tilde{\rho}(\tilde{z}) \leq 1.6$ so that the fluctuations are an order of magnitude smaller than the background density.

\subsection{Examples of Initial conditions}\label{sec:analysis_of_initial_conditions}
In Figure \ref{fig:my-label5} we present typical spatial profiles of the initial number difference field $\Tilde{\rho}$ (upper row) and phase difference field $\tilde{\phi}$ (lower row). Each profile provides the initial conditions for a single classical field trajectory and is obtained by summing up thermally activated phonons (Fourier modes) using the Tomonaga-Luttinger model.

The different columns show the effect of changing temperature $T$ or Luttinger parameter $K$. As expected, when $T$ is increased the fluctuations in both  $\Tilde{\rho}$ and $\tilde{\phi}$ increase. By contrast, if $K$ is increased the maximum magnitude and jaggedness of $\Tilde{\rho}$ 
increases but the jaggedness of $\tilde{\phi}$ decreases. Referring to Eq.\ (\ref{eq:coefficientlist}) we can see that this is because the coefficient multiplying the density fluctuation term in the Hamiltonian is $\Gamma= \pi / 2K$ which decreases as $K$ increases leading to increased variance of $\varrho_k$ modes according to Eq.\ (\ref{eq:sigmarhoplus}). The phase fluctuation term shows the opposite behavior because its coefficient in the Hamiltonian (which only appears as the spatial gradient of $\tilde{\phi}$) is $\epsilon=K/2 \pi$ which increases as $K$ increases and this reduces the variance of the $\varphi_{k}$ modes according to Eq.\ (\ref{eq:sigmaphiplus}), thereby  making the $\tilde{\phi}$ profiles smoother.



\section{Numerical Simulations of the Dynamics}\label{sec:caustics_in_sine_gordon}

In this section we explore the dynamics following a $J$-quench. Our approach is inspired by the TWA where multiple classical field configurations are propagated in time using the classical equations of motion, although in our case the initial conditions are sampled from a thermal distribution as described in Section \ref{sec:IC1} rather than a quantum distribution as in the standard TWA.

$J$-quench dynamics have previously been explored for the simpler case of a two-mode zero temperature bosonic Josephson junction where  it was found that caustics dominate the number and phase difference probability distributions \cite{DuncanSirPRL,Mumford2017,Mumford2019}. In the two-mode case it is possible to compute the \textit{exact quantum dynamics} for some thousands of particles and compare them against the TWA. The results (see Figure 1 in \cite{DuncanSirPRL}) show good qualitative agreement and give us confidence that the TWA can capture the main features of the quantum dynamics. Furthermore, the inevitable presence of decoherence due to the environment will tend to reduce the quantum dynamics to their classical limit (this has been investigated in the two-mode case for a $J$-quench in \cite{Goldberg2019}) increasing the relevance of semiclassical calculations.
In the present work we are interested in whether the phonons along the long axis  disrupt or sustain these caustics. We will start by reproducing the caustics presented in Ref.~\onlinecite{DuncanSirPRL} for the two-mode case and then add in the longitudinal modes after that.



\begin{figure*}[th]
\centering
\begin{tabular}{ccc}
     {\includegraphics[scale = 0.37]{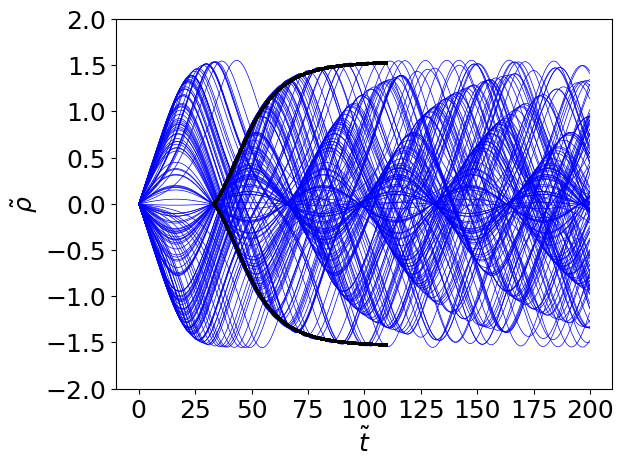}}
     {\includegraphics[scale = 0.37]{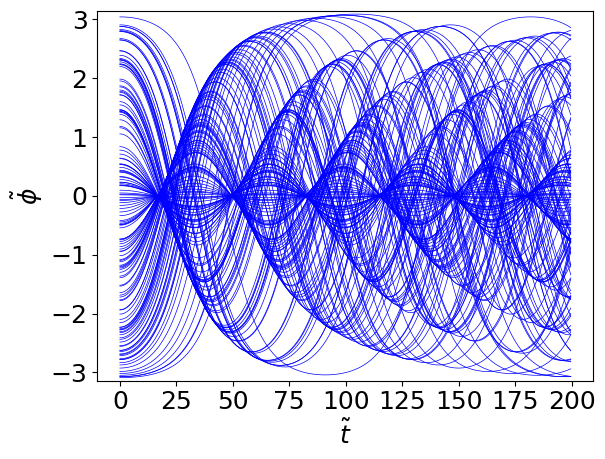}}
     {\includegraphics[scale = 0.37]{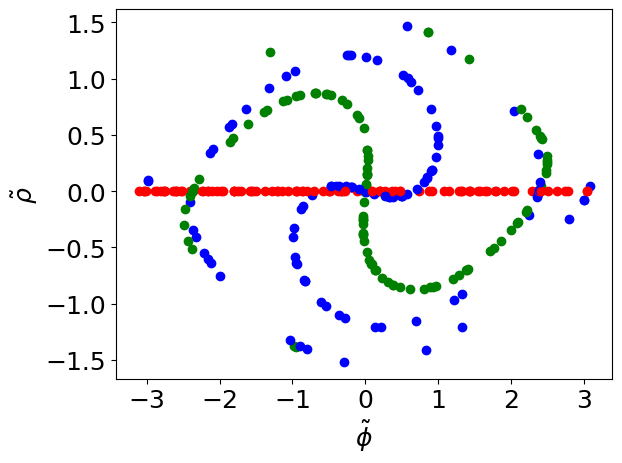}}\\
\end{tabular}
\caption{\label{fig:0D_dynamics} Dynamics in the two mode approximation governed by the Josephson equations given in Eq.\ (\ref{eq:eomSGplus0D}) of the number difference $\tilde{\rho}$ (left), phase difference $\tilde{\phi}$ (middle), and phase space distribution (right) following a quench from $J=0$ to $J=30$ Hz. The other parameter values are given in Table \ref{parameter_table}. Each panel contains 150 trajectories: each trajectory starts with $\tilde{\rho}=0$ at time $\tilde{t}=0$ but has an initial phase randomly sampled from $[-\pi,\pi)$. Both number and phase difference variables display a series of cusp shaped caustics given by the envelopes of families of trajectories; to guide the eye we have outlined the first cusp caustic in the $\tilde{\rho}$ variable with a black curve.  In the right panel three different time slices of the results are plotted in  phase space ($\Tilde{\rho}$ versus $\Tilde{\phi}$). Each dot corresponds to a different initial condition (trajectory) and the colors indicate the time: $\tilde{t}$=0 (red), $\tilde{t}$=50 (green), $\tilde{t}$=100 (blue). During time evolution the initial horizontal line winds into a whorl and the caustics in the $\Tilde{\rho}$  and  $\Tilde{\phi}$ plots  occur due to horizontal and vertical segments of a whorl, respectively.}
\end{figure*}



\subsection{Numerical Methods}

The initial conditions are generated via random sampling from Gaussian distributions. We then evolve the equations of motion (Eq.\ \ref{eq:eomSGplusdimensionless} 
for the case of the full SG+ model) using a Runge-Kutta solver with a user-defined time step \cite{rackauckas2017differentialequations}. The endpoints of our system are treated by imposing periodic boundary conditions. In  Appendix \ref{sec:benchmarking} we demonstrate the numerical convergence of the solver by varying the temporal and spatial steps by tracking the time evolution of the total energy (hamiltonian) which should be a constant of the motion and obtain the fiducial time and space resolution for all our calculations.

\begin{figure*}[ht]
\centering
 \includegraphics[width=0.65\columnwidth]{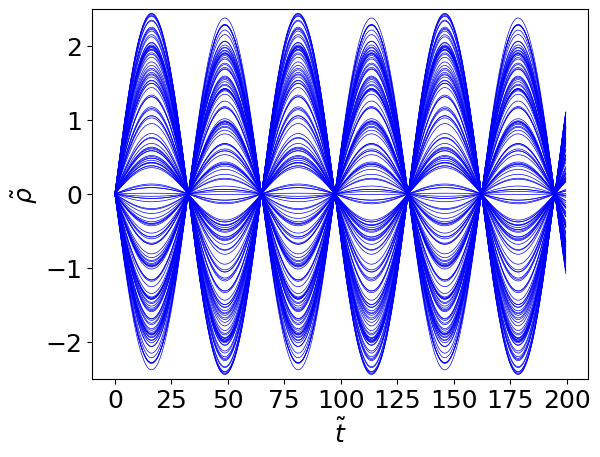}
  \includegraphics[width=0.65\columnwidth]{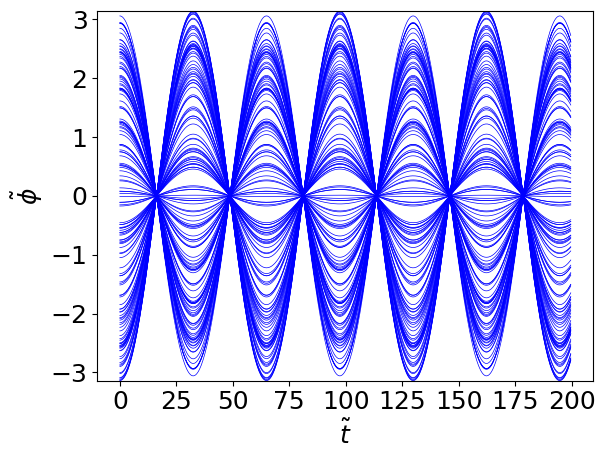}
 \includegraphics[width=0.65\columnwidth]{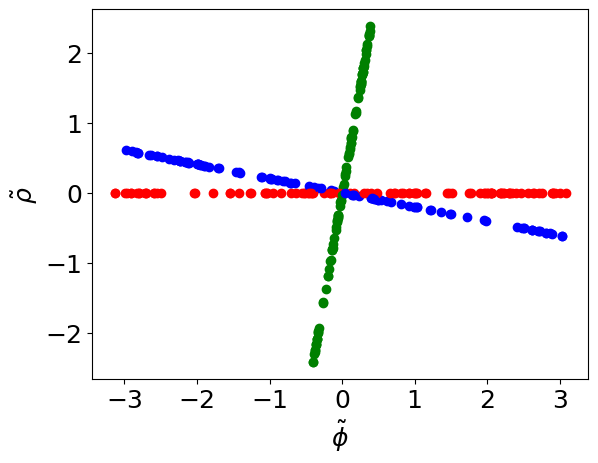}
\caption{\label{fig:0D_dynamicslinear} Dynamics in the linearized version of the two-mode approximation [Eq.\ (\ref{eq:eomSGplus0Dlinear})] of the number difference $\tilde{\rho}$  (left), phase difference $\tilde{\phi}$ (middle), and the phase space distribution (right)  following a quench from $J=0$ to $J=30$ Hz. Like in Figure \ref{fig:0D_dynamics}, there are 150 trajectories shown in each panel corresponding to different values of the initial value of $\tilde{\phi}$. However, in this linearized case we obtain a series of perfect focus points (revivals of the initial state). This is because linearization  gives rise to rigid rotation in phase space without whorls. Unlike the extended cusp caustics seen in Figure \ref{fig:0D_dynamics} (which will be qualitatively robust to details of the nonlinearity) perfect focus points are nongeneric because they are unstable to perturbations such as the effects of nonlinearity. All parameter values and color labels are the same as Figure \ref{fig:0D_dynamics}. }
\end{figure*}




\begin{figure*}[t]
\begin{tabular}{ccc}
     {\includegraphics[scale = 0.37]{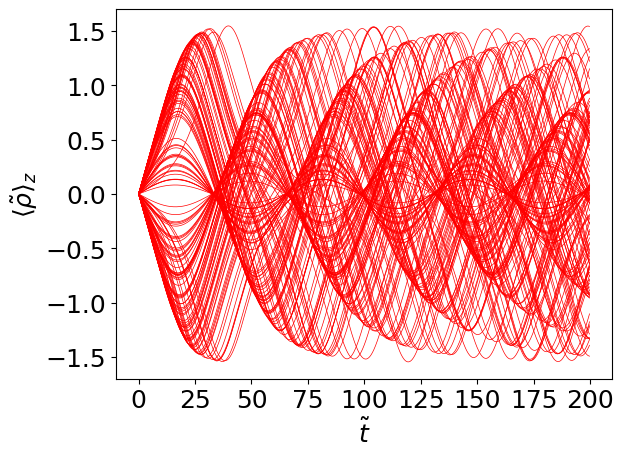}}
     {\includegraphics[scale = 0.37]{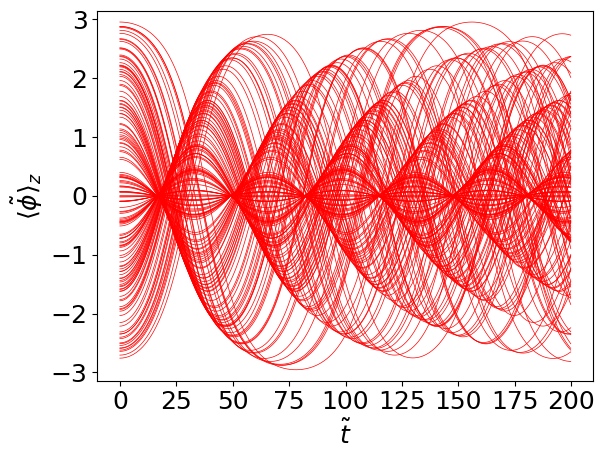}}
     {\includegraphics[scale = 0.37]{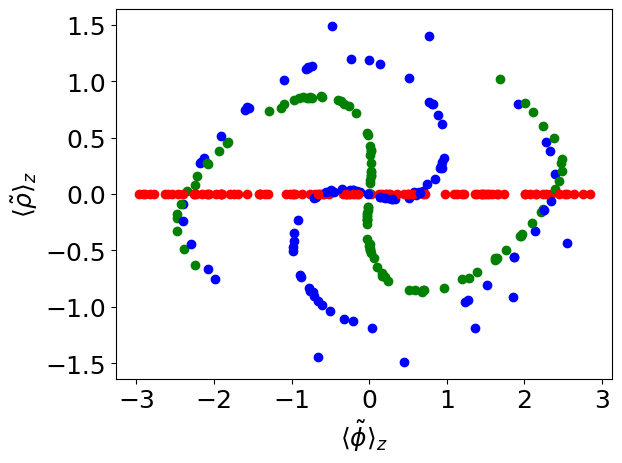}} 
\end{tabular}
\caption{\label{fig:nonlinear1D_dynamics}
Dynamics in the SG+ model of the spatially averaged number difference $\langle \tilde{\rho} \rangle_{z}$ (left), phase difference $\langle \tilde{\phi} \rangle_{z}$ (middle), and phase space distribution (right)  following a quench from $J=0$ to $J=30$ Hz. Each panel contains 150 trajectories which are solutions of Eq.\ (\ref{eq:eomSGplusdimensionless}).  The initial conditions are randomly sampled thermal phonons with the same parameter values as those shown in the top row of Figure  \ref{fig:my-label5} and described in Table \ref{parameter_table}. In particular, the number of numerical lattice points is $N_L=50$ separated by a grid spacing of $a=0.36 \ \mu$m, and the temperature is $T=2$ nK. The healing length is $\xi_h=0.24 \ \mu$m, the spin healing length is $\xi_s=2.5 \ \mu$m and the phase coherence length is $\lambda_T=380 \ \mu$m.  The different colors on the phase space plot correspond to the same time slices as in the previous phase space plots.}
\end{figure*}

\subsection{Special case: two-mode approximation} \label{subsec:0D}

In the two-mode approximation only a single mode in each well is taken into account.  This description is relevant to the SG/SG+ model in the limit where the entire length of each quasicondensate is perfectly synchronized so that the fields $\tilde{\rho}(\tilde{z})$ and $\tilde{\phi}(\tilde{z})$ do not depend on $\tilde{z}$. In this case the spatial derivative terms vanish and the equations of motion in Eq.\ (\ref{eq:eomSGplusdimensionless}) reduce to 
\begin{equation}
    \frac{d \tilde{\phi}}{d \tilde{t}} =2 \Gamma \tilde{\rho} \quad , \quad 
    \frac{d \tilde{\rho}}{d \tilde{t}} =  - 2 \mathcal{J} \sin \tilde{\phi} \ .
\label{eq:eomSGplus0D}
\end{equation}
These are the standard Josephson equations  of motion and also correspond to those of a classical pendulum \cite{pitaevskiistringaribook}. Such synchronization can occur at very low temperatures or when the coefficients $\epsilon$ and $\Gamma$ are large enough that they suppress spatial fluctuations in the initial conditions.

In Figure \ref{fig:0D_dynamics} we display
the post-quench dynamics in the two-mode approximation. The left hand and central panels show the  time dependence of $150$ independent solutions
of Eq.\ (\ref{eq:eomSGplus0D}) which give the trajectories for the number difference and phase difference, respectively. Note that in this paper we use the color blue for trajectories calculated within the two mode approximation and reserve red for the trajectories of the full many mode model. In accordance with our assumption that the two wells start with an equal number of atoms, each solution starts with $\tilde{\rho}=0$.  And as discussed in Section \ref{subsec:IC2}, the initial value of $\tilde{\phi}$ is randomly chosen from the range $[-\pi,\pi)$ because the two condensates are independent before the $J$-quench. 

The most striking feature of Figure \ref{fig:0D_dynamics} is the series of cusp-shaped caustics that form in both variables. In order to guide eye, we have have outlined the first cusp caustic in the number difference variable using a black curve (the calculation for this curve is given in Appendix \ref{app:envelope}).  Like in optics, caustics are regions of high intensity formed by the envelopes of families of rays (trajectories). Each caustic is born at the centre of the distribution at the tip of a cusp before spreading out in two arms that move towards the edges of the distribution. The fact that they are cusp shaped is in agreement with the prediction of catastrophe theory that in two dimensions the only structurally stable and hence generic singularities are cusps.

Each trajectory represents a single experimental run. The idea behind the TWA is that the number of trajectories reaching a point $\tilde{\rho}$ at time $\tilde{t}$ is proportional to the probability that a measurement of the true quantum system would yield that value of $\tilde{\rho}$. An equivalent interpretation holds for the $\tilde{\phi}$ trajectories. The caustics have the highest probability density and hence give the values most likely to be observed. Of course, if we only consider the average values of $\tilde{\rho}$ or $\tilde{\phi}$ we would get zero in both cases due to the symmetry of the distributions and hence miss the caustics. Many experimental runs must be performed in order to obtain the probability distribution where these patterns live.

The mechanism underlying caustics can be understood from a phase space perspective, as shown in the right hand panel of Figure \ref{fig:0D_dynamics}. Each dot gives the number and phase difference at a particular time for a different initial condition. The red dots are the initial values which lie in a horizontal line because at $\tilde{t}=0$ all trajectories have  $\tilde{\rho}=0$. As time evolves the dots rotate around the origin: the green and blue dots show two successively later times. However, the nonlinearity of the Josephson equations means dots further from the origin rotate more slowly and this leads to the formation of a spiral or whorl. At places where the whorl has a vertical segment a range of different solutions all have the same value of $\tilde{\phi}$ and this stationarity of the distribution with respect to changes in the initial conditions is what generates a caustic, in this case a $\tilde{\phi}$-caustic. Conversely, horizontal segments give rise to $\tilde{\rho}$-caustics. 

In the absence of nonlinearity the equations reduce to those of a harmonic oscillator 
\begin{equation}
    \frac{d \tilde{\phi}}{d \tilde{t}}  =2 \Gamma \tilde{\rho} \quad , \quad
    \frac{d \tilde{\rho}}{d \tilde{t}} =  - 2 \mathcal{J}  \tilde{\phi} 
\label{eq:eomSGplus0Dlinear}
\end{equation}
giving rise to rigid rotation in phase space and the formation of perfect focal points in the number and phase difference variables, as shown in Figure \ref{fig:0D_dynamicslinear}. However, these perfect revivals of the initial state are not stable: any nonlinearity will cause the focal points to evolve into the extended cusp caustics shown in Figure \ref{fig:0D_dynamics}.

The frequency of the linearized motion is known in Josephson junction terminology as the plasma frequency. In our notation it reads
\begin{equation}
\label{eq:plasma_frequency}
    \omega_{p}=\sqrt{4 \Gamma \mathcal{J}} 
\end{equation}
and the period of the motion is therefore given by $2 \pi / \omega_{p}$. For the case shown in Figure \ref{fig:0D_dynamicslinear} we have $\Gamma=0.063$ and $\mathcal{J}=0.037$ giving a period $\approx 65$. In fact, the tips of the cusps in the nonlinear case also occur with this period since they are formed from small amplitude trajectories that only experience the quadratic bottom of the cosine potential.

\subsection{General case: many-mode SG+ model}\label{subsec:1D} 

Simulations of the full SG+ model are shown in Figure \ref{fig:nonlinear1D_dynamics}, which represents one of the main results of this paper. The trajectories in the left panel give the spatially averaged number difference $\langle \tilde{\rho}(\tilde{t}) \rangle_{z}$ as a function of time obtained by solving the equations of motion given in Eq.\ (\ref{eq:eomSGplusdimensionless}) for the many-mode system  and then averaging over its length. The trajectories in the middle panel of Figure  \ref{fig:nonlinear1D_dynamics}  give the equivalent spatial average of the phase difference $\langle \tilde{\phi}(\tilde{t}) \rangle_{z}$, and the right-hand panel is the phase space picture. Each trajectory is evolved from a single randomly sampled field configuration (describing thermally activated phonons) such as those shown in the top row of Figure \ref{fig:my-label5} and for the parameters given in Table \ref{parameter_table}.   We observe that despite the inclusion of longitudinal modes and the randomness of the initial conditions, the caustics survive and are quite similar to those of the two-mode approximation shown in Figure \ref{fig:0D_dynamics}. 
This suggests that caustics are a generic feature of many particle dynamics following quenches, at least for systems whose underlying physics is based on coupled nonlinear oscillators. Each oscillator starts with a random phase and a noisy momentum but the quench acts so as to give all the oscillators a momentum kick at the same time $\tilde{t}=0$ leading to an initial partial synchronization. As the system evolves in time after the kick the different periods of nonlinear oscillators leads to cusp catastrophes in the distribution of trajectories. If we had instead calculated only the \textit{expectation values} of the number and phase differences then this underlying structure would not have been visible because it lives in the probability distribution rather than the mean values.

\begin{figure}[t]
\begin{tabular}{c}
     {\includegraphics[scale = 0.52]{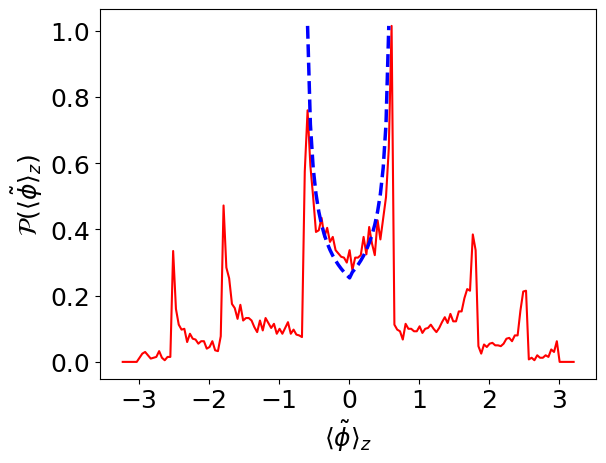}}\\
     {\includegraphics[scale = 0.52]{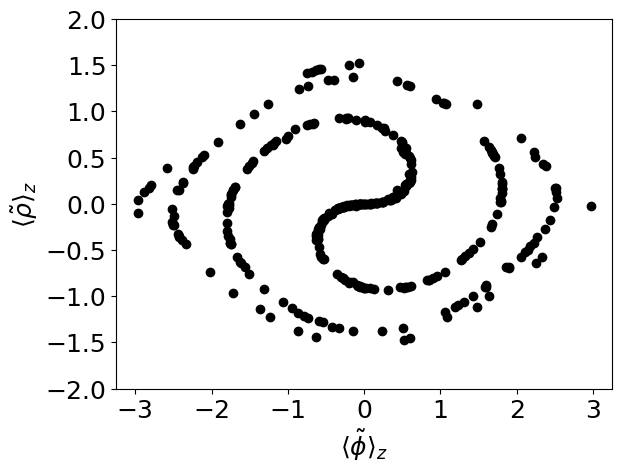}}
\end{tabular}
\caption{\label{fig:nonlinear_density_of_trajectories}  Top: Normalized probability distribution (red curve) as a function of $\langle \tilde{\phi} \rangle_{z}$ at time $\tilde{t}=162$ obtained from the density of trajectories with a bin width $d \tilde{\phi} = 0.04$  for the SG+ model.  This corresponds to a slice at fixed time through the middle panel of Figure \ref{fig:nonlinear1D_dynamics}, although calculated using 10000 trajectories to improve the statistics and averaged over a short time window of $\Delta \tilde{t} = 1$ to remove rapid fluctuations.  Caustics appear as sharp peaks and are well fitted (blue dashed curves) by the inverse square root form given in Eq.\ (\ref{eq:sqrootsingularity}).  The satellite caustics at   $\langle \tilde{\phi} \rangle_{z} \approx \pm 1.8$, $\pm 2.5$, $\pm 2.9$ also have this shape but the fit is not shown to avoid obscuring the data. Bottom: The phase space distribution for the same time slice. Each vertical segment of the whorl lines up with a caustic in the probability distribution of the phase variable in the top panel.}
\end{figure}



A slice at fixed time through the probability distribution for the spatially averaged phase variable $\langle \tilde{\phi} \rangle_{z}$ is shown in the top panel in Figure  \ref{fig:nonlinear_density_of_trajectories}. This is obtained by sorting the $\langle \tilde{\phi} \rangle_{z}$ trajectories into bins each of which covers a small range of $\langle \tilde{\phi} \rangle_{z}$ and plotting the number of trajectories in each bin. The result is noisy due to the thermal fluctuations but the caustics are clearly visible as sharp peaks. These peaks display the characteristic `square root' divergence of fold caustics \cite{Nye_natural_focusing}
\begin{equation}
 \mathcal{P} (\langle \tilde{\phi} \rangle_{z})  \propto  \frac{1}{\sqrt{ \tilde{\phi}_c -  \langle \tilde{\phi}\rangle_{z}}}
 \label{eq:sqrootsingularity}
\end{equation}
where $\mathcal{P}(\langle \tilde{\phi} \rangle_{z})$ is the  probability density and  $\tilde{\phi}_{c}$ is the location of the caustic. The blue dashed lines in Figure  \ref{fig:nonlinear_density_of_trajectories} are fits of Eq.\ (\ref{eq:sqrootsingularity}) to the numerical data and we see that the agreement is good. 
Although the height of the singularities predicted by Eq.\ (\ref{eq:sqrootsingularity}) is infinite at the caustic, this function is integrable so that a probability distribution with caustics is still normalizable (of course, the peaks in the numerical data are of finite height because the number of trajectories is finite). The positions of the caustics can be matched exactly to the vertical segments of the phase space distribution shown in the bottom panel of Figure \ref{fig:nonlinear_density_of_trajectories}. The horizontal segments give rise to a similar pattern of caustics in the probability distribution for the number difference variable  (not shown in this figure).

\subsection{Effect of temperature on the caustics} \label{subsec:temp}

\begin{figure*}[t]
\centering
\begin{tabular}{ccc}
     {\includegraphics[scale = 0.37]{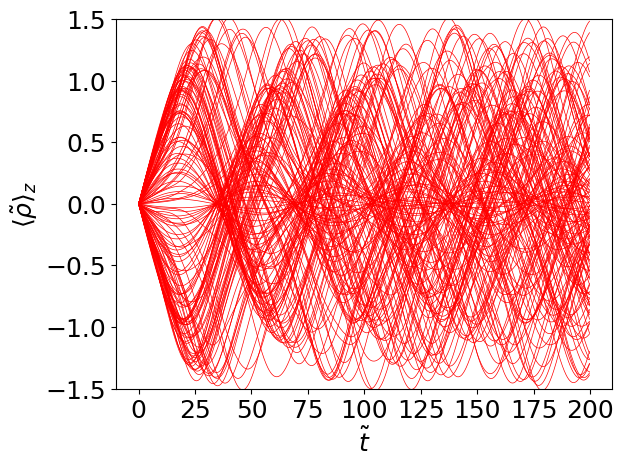}}
     {\includegraphics[scale = 0.37]{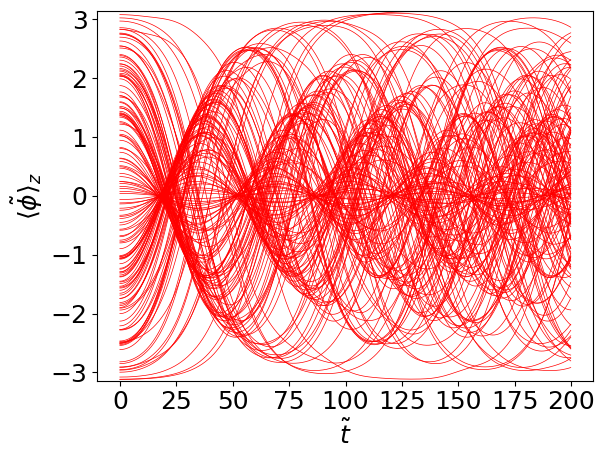}}
     {\includegraphics[scale = 0.37]{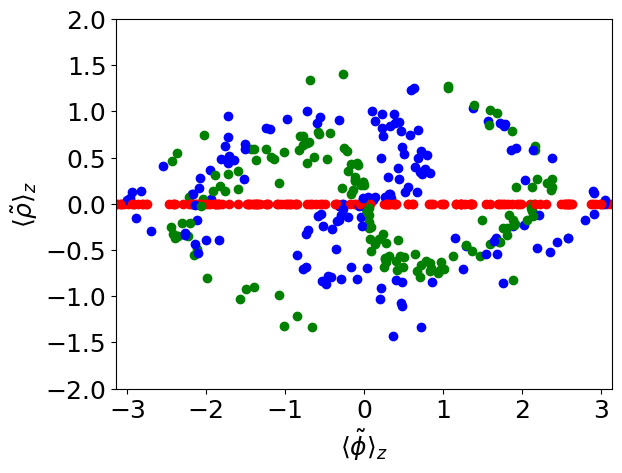}}
\end{tabular}
\caption{\label{fig:temperature} Effect of temperature on the spatially averaged number difference $\langle \tilde{\rho} \rangle_{z}$ (left), phase difference $\langle \tilde{\phi} \rangle_{z}$ (middle), and phase space (right) for the SG+ model. Each panel contains 150 trajectories with parameters and color schemes the same as those used for Figure \ref{fig:nonlinear1D_dynamics} except the temperature here is T = 50 nK giving a thermal phase coherence length $\lambda_T$ = 15.2 $\mu m$.} 
\end{figure*}

\begin{figure}
\centering
\begin{tabular}{cc}
     {\includegraphics[scale = 0.52]{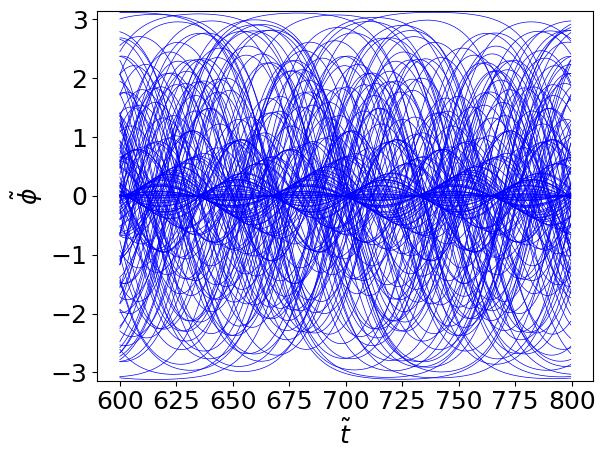}}\\
     {\includegraphics[scale = 0.52]{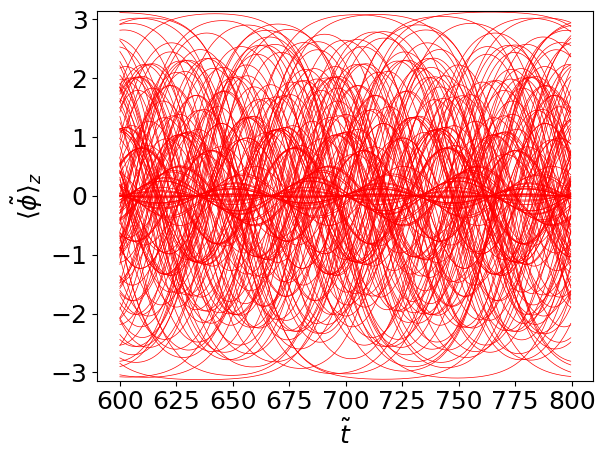}}\\
     {\includegraphics[scale = 0.52]{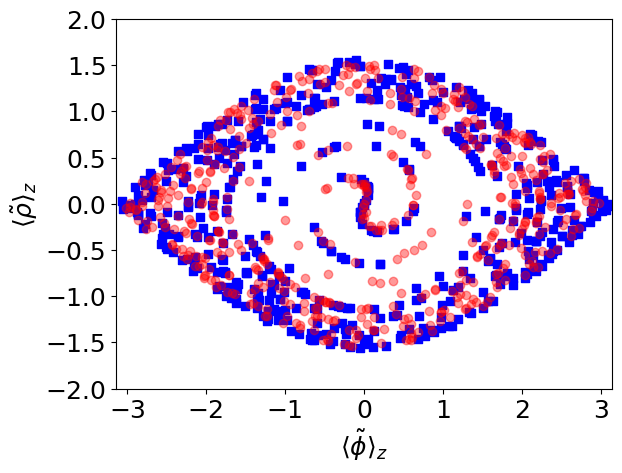}}
\end{tabular}
\caption{\label{fig:double_derivative} Effect of dispersion: comparison of the long-time behavior of the phase difference in the two-mode approximation (top), SG+ model (middle), and their respective phase space behaviours at $\tilde{t}=800$ (bottom). The top two panels contain 150 different runs with the same parameters as those used in Figure  \ref{fig:nonlinear1D_dynamics} except that $\epsilon$ has been artificially multiplied by 10  in the middle panel. Caustics are visible in the upper panel but less so in the middle panel. In the bottom panel the blue squares (two-mode approximation) show better-defined whorls than the red circles (SG+ model) which are more randomly dispersed near the edges of the eye shaped boundary.}
\end{figure}

In Figure \ref{fig:temperature} we investigate the effect of finite temperature on the formation of caustics in the SG+ model by using the same parameters as Figure \ref{fig:nonlinear1D_dynamics} except that we raise the temperature in the initial conditions from 2 nK to 50 nK. Comparing the plots in Figures \ref{fig:nonlinear1D_dynamics} and \ref{fig:temperature} we see that although the first few caustics after the quench are still visible in the latter, a higher temperature seems to wash out the caustics after that. In the bottom panel of Figure \ref{fig:temperature} we see why this is: the whorl in phase space becomes blurred by temperature induced fluctuations.  However, the magnitude of this effect is dependent on the other parameters. As will be explained in Sec.\ \ref{sec:effect_of_J_quench}, increasing the value of $J$ can make the caustics more prominent again by putting more energy into the post-quench dynamics in comparison to the thermal energy. Nevertheless, the persistence of caustics even at higher temperatures illustrates their key property of structural stability against perturbations.

\begin{figure*}[ht]
\centering
\begin{tabular}{ccc}
     {\includegraphics[scale = 0.37]{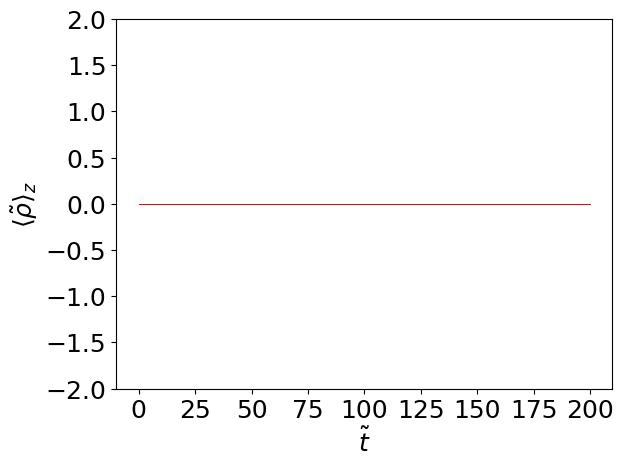}}
     {\includegraphics[scale = 0.37]{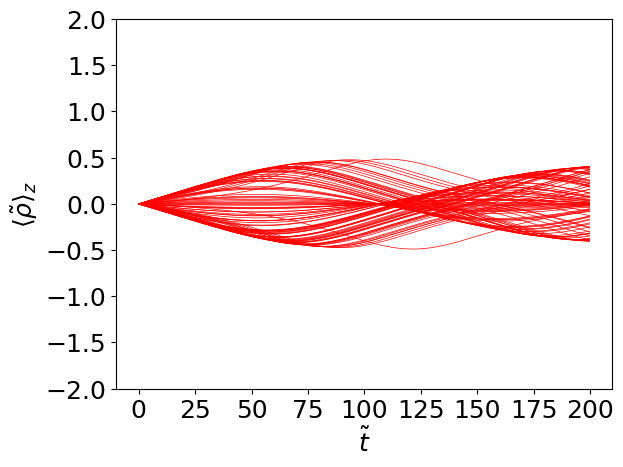}}
     {\includegraphics[scale = 0.37]{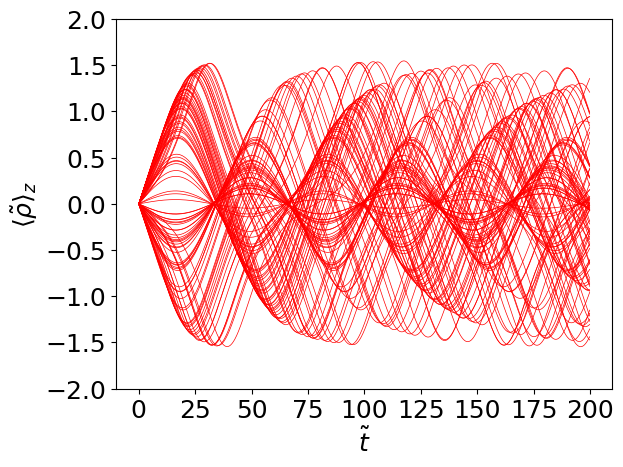}}\\
     {\includegraphics[scale = 0.37]{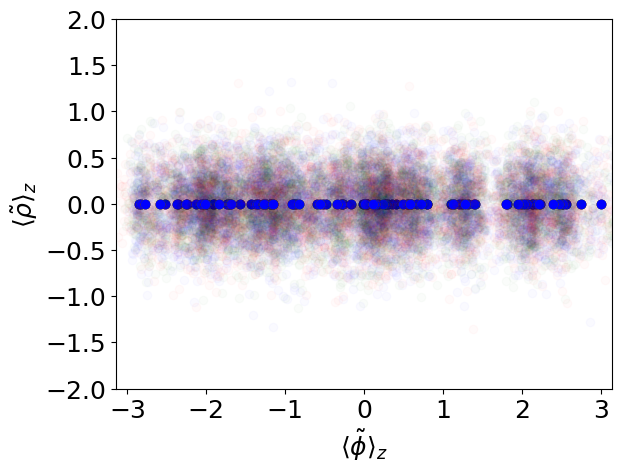}}
     {\includegraphics[scale = 0.37]{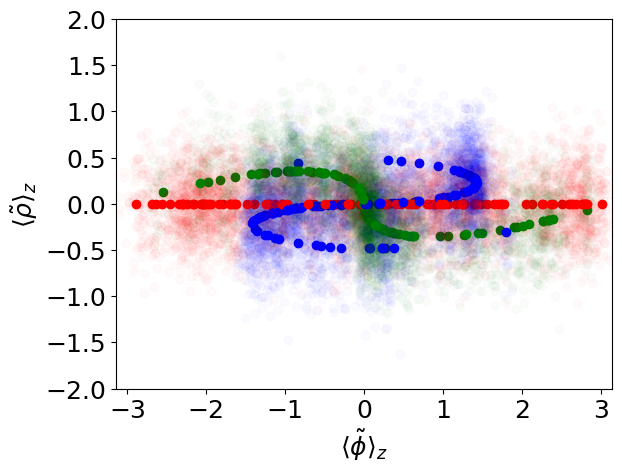}}
     {\includegraphics[scale = 0.37]{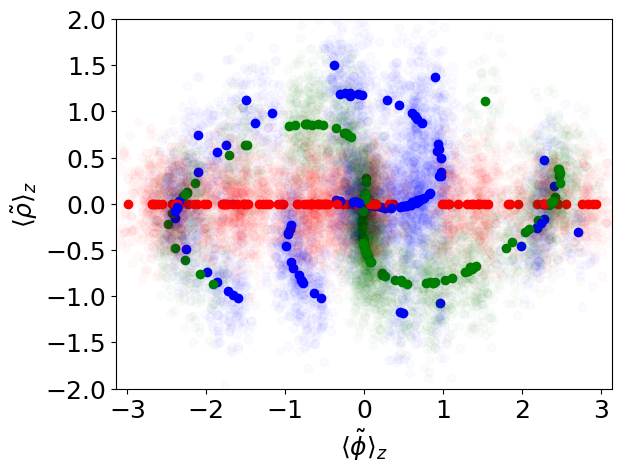}}
\end{tabular}
\caption{\label{fig:dynamics_no_quench} Effect of quench strength $J$ for $J=0$ Hz, 3 Hz, and 30 Hz (from left to right). The top row shows the dynamics of $\langle \tilde{\rho} \rangle_z$ with initial conditions sampled in the same way as in Figure \ref{fig:nonlinear1D_dynamics}.  The bottom row plots the corresponding phase space distributions. Like in previous figures, the different colors give different time instants: $\tilde{t}$=0 (red), $\tilde{t}$=50 (green), $\tilde{t}$=100 (blue). The dots with intense colors are the spatially averaged values. We have also included the raw data (without spatial averaging) as faint dots. This gives an idea of the size of the statistical fluctuations due to the thermal initial conditions and is the same for all values of $J$. In the left column there is no coupling between the two quasicondensates and hence no time evolution of the spatially averaged data (the intense red, green, and blue dots sit on top of each other) although there can be evolution of unaveraged data due to intrawell dynamics, i.e.\ without the $J$ term in Eq.\ \eqref{eq:EOM1}.  As we increase the magnitude of $J$ time evolution leads to whorls with a greater vertical extent because more energy can be extracted from the cosine potential in Eq.\ (\ref{eq:HSGplus_dimensionless}) giving larger values of $\langle \tilde{\rho} \rangle_{z}^{\mathrm{max}}$.}
\end{figure*}


\subsection{Effect of dispersion on the caustics} \label{subsec:spatial_term}

The double derivative terms in the SG+ equations of motion given in  Eq.\  (\ref{eq:eomSGplusdimensionless}) are responsible for transmitting wave disturbances along the longitudinal axis and are not present in the simpler two-mode case discussed in Section \ref{subsec:0D}. Initial thermal fluctuations in the SG+ model will therefore disperse in $z$ over time and it is interesting to see what difference this makes to the caustics; comparison of Figures \ref{fig:0D_dynamics} and \ref{fig:nonlinear1D_dynamics} suggests it makes little difference to spatially averaged variables. However, this observation is for only one choice of the parameters $\epsilon$ and $\Gamma$ that govern the size of the derivative terms and also for relatively short times. In particular, in Figure  \ref{fig:nonlinear1D_dynamics} the  parameters are $\epsilon \approx 4$  and $\Gamma \approx 0.06$ which were chosen to match experimental values \cite{Gring_and_Schmeidmayer_relaxation_paper,Schmeidmayer_double_light_cone_paper,Schweigler_2017,Local_emergence_and_correlation_paper,Rauer2018,Pigneur2018}.  In Figure \ref{fig:double_derivative} we compare the long time dynamics of the two-mode approximation and the SG+ model for the case where $\epsilon$ in the SG+ model has been artificially increased by a factor of 10 (without changing any other parameters), thereby increasing the effect of spatial dispersion. Apart from this change, the initial conditions and $J$-quench are similar to those used in Figure  \ref{fig:nonlinear1D_dynamics}. Note that we only use this increased value of $\epsilon$ for the time propagation and not for the generation of the thermal initial conditions. This avoids changing the starting phase fluctuations from those used in Figure \ref{fig:nonlinear1D_dynamics} which would otherwise be energetically suppressed and would also lead to significantly different dynamics but is not the comparison we would like to make here. From Figure \ref{fig:double_derivative} we see that the strong coupling of neighboring `pendula' does wash out the caustics at long times in comparison to the dispersionless two-mode case, although examining the phase space distribution  in the bottom panel the inner part of the whorl  is still visible which further underlines the robustness of caustics. Note that in the bottom panel (and only this panel) we have biased the sampling of SG+ trajectories (red circles) so that a greater fraction of them are near the edges than would otherwise occur naturally. This is simply to increase the visibility near the edges where we see the SG+ trajectories are  randomly dispersed. The long time behavior of both the SG+ model and the two-mode approximation will be further analyzed in Section \ref{sec:importance}.

\subsection{Effect of $J$ on the caustics}
\label{sec:effect_of_J_quench}

Another parameter that affects the dynamics is the tunnel coupling strength $J$ [or its dimensionless version $\mathcal{J}$ which is defined in Eq.\ (\ref{eq:coefficientlist})] that becomes non-zero after the quench. The quench itself creates a strongly nonequilibrium phase difference where all values of $\tilde{\phi}$ are equally probable independently of the value of $J$ by virtue of the fact that before the quench there is no phase coherence between the two quasicondensates. However, $J$ does control the post-quench dynamics. One way it does this is via the frequency of the Josephson oscillations. The cusps occur with a frequency given by the plasma frequency in Eq.\ (\ref{eq:plasma_frequency}) which goes as $\sqrt{J}$.

 In Figure \ref{fig:dynamics_no_quench} we examine the effect of quenching to different $J$ values, with the value of $J$ increasing from left to right. We can see the expected increase in frequency. The amplitude of the motion also increases with $J$ because immediately after the quench each trajectory finds itself at a random point on the cosine potential energy surface whose depth between valley top and valley bottom is $2 \mathcal{J}$. The initial potential energy of a field configuration is therefore $-2\mathcal{J}\langle \cos  \tilde{\phi}_{0} \rangle_{z}$, where $\tilde{\phi}_{0}$ is the phase field $\tilde{\phi}(\tilde{z},\tilde{t})$ at the initial time.  This configuration evolves under the full Hamiltonian and gives rise to oscillations about the potential minimum. The upper row in Figure \ref{fig:dynamics_no_quench} plots the spatially averaged number difference and according to Eq.\ (\ref{eq:HSGplus_dimensionless}) the maximum amplitude this can have is
 \begin{equation}
 \label{eq:rhomax}
     \langle \tilde{\rho} \rangle_{z}^{\mathrm{max}} = \sqrt{\frac{2 \mathcal{J}(1- \langle \cos \tilde{\phi}_{0} \rangle_{z})}{\Gamma}}
\end{equation}     
 where we have ignored the effects of spatial coupling (second order derivative terms). Thus, $\langle \tilde{\rho} \rangle_{z}^{\mathrm{max}}$ also scales as $\sqrt{J}$, and this is in correspondence with Figure \ref{fig:dynamics_no_quench}.
 
The lower row of Figure \ref{fig:dynamics_no_quench} shows the behavior in phase space. In these figures we have also included the unaveraged data, i.e.\ the $\tilde{\rho}$ and $\tilde{\phi}$ values of each grid point at the three selected times. This gives a sense of the size of the statistical fluctuations due to the spatial degrees of freedom. In the left hand column $J$ remains zero for all time and the only dynamics that can occur is along the long-axis of each quasicondensate individually. The middle and right hand panels, which have $J=3$ and $J=30$ Hz, respectively, have the same initial statistical fluctuations as the left hand one because, as mentioned above, the initial distribution is set by the pre-quench thermal fluctuations in the two quasicondensates and is independent of $J$. However, as time evolves the effects of $J$ described by Eq.\ (\ref{eq:rhomax}) become apparent because larger $J$ allows a greater value of  $\langle \tilde{\rho} \rangle_{z}^{\mathrm{max}}$ and this stretches the distribution along the vertical direction in comparison to a smaller value of $J$. For a whorl to become apparent $\langle \tilde{\rho} \rangle_{z}^{\mathrm{max}}$ should at least exceed the width of the statistical fluctuations and becomes better and better defined as $J$ is increased.

\section{Universality and caustics} \label{sec:importance}

We have already discussed the relationship between nonlinearity and caustics in the preceding section. As 
motivated earlier, and
expounded in Refs.~\onlinecite{DuncanSirPRL,Mumford2017,Plestid2018,Mumford2019,Kirkby2022}, caustics also have implications for the universal dynamics of quantum systems. We explore a few of these effects in this section.

\begin{figure}[t]
\begin{tabular}{c}
     {\includegraphics[scale = 0.52]{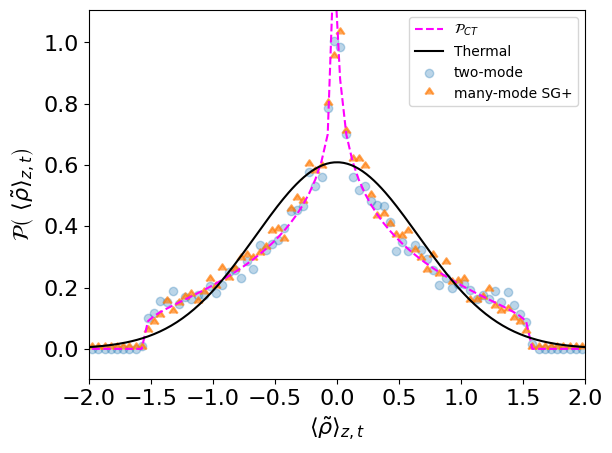}}\\
\end{tabular}
\caption{\label{fig:probability_intensity} The long time probability distribution for the number difference $\tilde{\rho}$. The data points are from the different nonlinear models considered in this paper averaged over the spatial coordinate $z$ and also over a time window ranging from $\tilde{t}=800$ to $\tilde{t}=980$ to remove fluctuations. The pink dashed line is the circus tent distribution $\mathcal{P}_{\mathrm{CT}}$ given in Eq.\ (\ref{eq:circustenta}) and derived in Appendix \ref{app:longtime} under the assumption of ergodicity; the circus tent shape is due to the proliferation of caustics at long times and gives a good fit to the data.  The solid black curve is the thermal distribution $\mathcal{P}_{T}$
 with a temperature chosen so that the expectation value of the energy matches that provided by the quench.}
\end{figure}


\subsection{Long time distribution: the circus tent}

 The quench generates collective excitations that lead to caustics as shown in Figures \ref{fig:0D_dynamics} and \ref{fig:nonlinear1D_dynamics} for the two nonlinear models (two mode and SG+) discussed above. The caustics are born at the center of the probability distribution (in either the $ \tilde{\rho}$ or the $ \tilde{\phi}$ variable) at intervals of the plasma period and move out to the 
edges over time. Figure \ref{fig:nonlinear_density_of_trajectories} plots the probability distribution for the SG+ model as a function of $\langle \tilde{\phi} \rangle_{z}$ at an intermediate time where four pairs of fold caustics are discernible and shows how they diminish in strength but are still present as they move to the edges. The question then naturally arises as to what happens at long times $\tilde{t} \rightarrow \infty$ when the distribution comprises of a large number of caustics and whether it tends to a characteristic shape?  The answer is yes, and is shown in Figure \ref{fig:probability_intensity} which is made in the same way as Figure  \ref{fig:nonlinear_density_of_trajectories} but this time by calculating the density of $\langle \tilde{\rho} \rangle_{z}$ trajectories and averaging  over a time window extending between $\tilde{t} = 800$ and $\tilde{t} = 980$ in order to remove rapid fluctuations.
The probability distribution takes a shape reminiscent of a `circus tent' or `big top' and can be understood as follows. The strongest singularities present are the cusp tips born at the center of the distribution which leads to this being the highest point. Each cusp then splits into two fold arms (which according to catastrophe theory are lower singularities) that move outwards, reducing in height as they go, before accumulating at the edges where there is a sharp drop to zero. The position of the outer edge is set by the maximum energy that can be extracted from the quench and is given by Eq.\ \ref{eq:rhomax}.  

\begin{figure*}[ht]
\centering
\begin{tabular}{ccc}
     {\includegraphics[scale = 0.37]{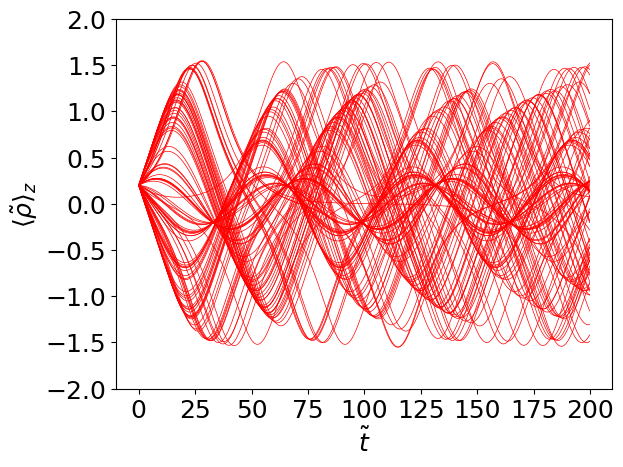}}
     {\includegraphics[scale = 0.37]{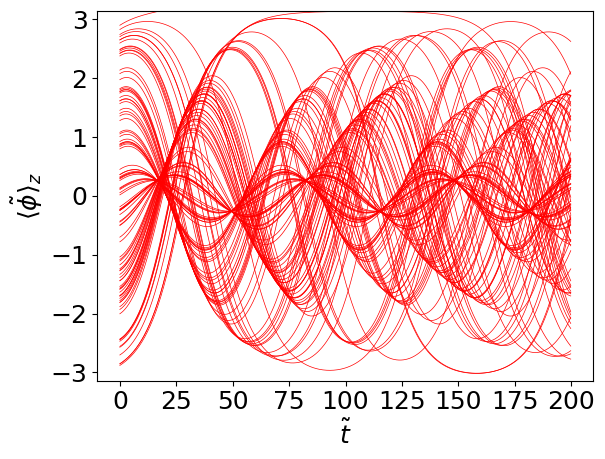}}
     {\includegraphics[scale = 0.37]{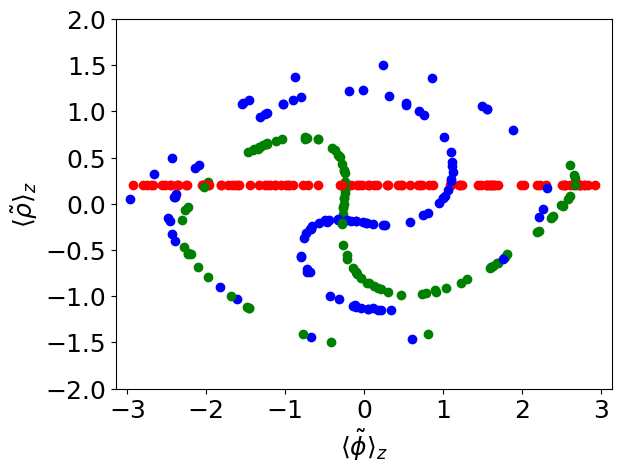}}
\end{tabular}
\caption{\label{fig:stability} Structural stability of caustics: here we investigate the effect of unbalanced densities on caustics  by tracking the same SG+ model dynamics as those shown in Figure \ref{fig:nonlinear1D_dynamics} except for an initial density imbalance of 0.1 in the background of $\tilde{\rho}$ at each point z. We see that the cusp caustics in the plots of $\langle \tilde{\rho} \rangle_{z}$ and $\langle \tilde{\phi} \rangle_{z}$ versus time are distorted but still maintain their basic structure. This is because the whorl in phase space is left intact despite having a displaced centre. Caustics are resilient against imperfections and perturbations and  we expect them to be present under realistic experimental conditions. }
\end{figure*}


An analytic expression for the circus tent distribution is given by the integral
\begin{equation}
\label{eq:circustenta}
   \mathcal{P}_{\mathrm{CT}}(\tilde{\rho})=\frac{1}{2 \pi B} \int_{\tilde{\rho}^2/B^2}^{1} \frac{U(m,\tilde{\rho})}{K(m)} \ dm
\end{equation}
where 
\begin{equation}
    U(m,\tilde{\rho})=\frac{1}{\sqrt{m(1-m)(m-\tilde{\rho}^2/B^2)(1+\tilde{\rho}^2/B^2-m)}},
    \end{equation}
$K(m)$ is the complete elliptic integral of the first kind, and $B=2 \sqrt{\mathcal{J}/\Gamma}$. This expression
is plotted in Figure \ref{fig:probability_intensity}
as the dashed line and is derived in Appendix \ref{app:longtime} under the assumption that at long times we can model the system by an ensemble of independent pendulua  where each pendulum is \textit{ergodic}. In other words, each pendulum obeys a microcanonical distribution where there is equal probability for it to be found anywhere on its energy shell. The nature of the $J$-quench is such that it leads to an ensemble with an equal probability for any starting angle (this is different to an equal probability for each energy due to the dependence of the density of states on angle).  As can be seen from Figure \ref{fig:probability_intensity}, $ \mathcal{P}_{\mathrm{CT}}(\tilde{\rho})$ gives a good fit to the numerical data generated by both the SG+ and two-mode models considered in this paper. 
For completeness, in Appendix \ref{app:longtime} we also give a plot (Fig.\ \ref{fig:long_time_phase}) of the long-time probability distribution for the phase difference $\tilde{\phi}$ and this also turns out to have a nonthermal circus tent-like shape.

In Figure  \ref{fig:probability_intensity} we also include the thermal probability distribution
\begin{equation}
\label{eq:totalproba}
    \mathcal{P}_{T}(\tilde{\rho}) = \frac{1}{Z} \int_{0}^{\infty} \mathcal{P}_{E}(\tilde{\rho}) \ e^{-E/T} D(E) \ dE
\end{equation}
describing an ensemble of pendula at thermal equilibrium at temperature $T$ where $\mathcal{P}_{E}(\tilde{\rho})$ is the probability distribution at fixed energy $E$, $D(E)$ is the density of states and
$Z$ is a normalizing factor. The details of our calculation of $\mathcal{P}_{T}(\tilde{\rho})$ are given in Appendix \ref{app:thermalpendulum}, where, for example, $\mathcal{P}_{E}(\tilde{\rho})$ is given in Eq.\ (\ref{eq:P_E}). The temperature of this  distribution is chosen such that the mean energy of the thermal distribution $\langle E \rangle_{T}$ is equal to the mean energy of the states excited by the quench. For a quench to $J=30$ Hz we show in Appendix \ref{app:thermalpendulum} that the effective temperature is 5.4 nK.

Clearly, the thermal distribution is very different to the circus tent distribution: the thermal distribution takes the form of a smooth gaussian with wings that extend beyond $\langle \tilde{\rho}\rangle_{z}^{\mathrm{max}}$ because the thermal Boltzmann factor allows for excitations with any energy (albeit with exponentially small probability) including those involving pendula undergoing rotation as well as libration, whereas the $J$-quench only excites librational motion. The probability distribution for a thermal pendulum is in fact quite delicate to compute because of the singularity in the density of states between libration and rotation but the combined result is smooth; see Appendix \ref{app:thermalpendulum} for more details.

\subsection{Structural stability of caustics}
\label{sec:structuralstability}

The defining characteristic of the singularities described by catastrophe theory is structural stability against perturbations and this ensures that they occur generically. The same is not true of isolated singularities as can be seen by comparing Figures \ref{fig:0D_dynamics} and \ref{fig:0D_dynamicslinear} where it is shown that point foci do not survive the introduction of nonlinearity. In two dimensions cusps are the unique structurally stable catastrophe and in Sections \ref{subsec:temp}  and \ref{subsec:spatial_term} we saw that cusp-shaped caustics are indeed stable against thermal fluctuations and the effects of dispersion.  However, thus far we have imposed the symmetrical starting condition that the initial number difference between the two quasicondensates is zero. One may therefore wonder whether the caustics we see are a consequence of this symmetry.  To check that this is not the case we show in Figure \ref{fig:stability} the dynamics for the case where the initial background density $n_{1D}$ in the two quasicondensates differs by $10 \%$.   We see that although the caustics in both $\langle \tilde{\rho} \rangle_{z}$ and  $\langle \tilde{\phi} \rangle_{z}$ are distorted they maintain their basic cusp shape. Furthermore, the phase space whorls still occur and this guarantees the existence of caustics.

\subsection{Coherence factor and relaxation towards equilibrium}

Cold atom experiments have the ability to measure correlation functions in nonequilibrium many-body states \cite{cheneau_lightcone_2012,Schweigler_2017,oberthaler2017,de_Nova2019}. As a simple example let us consider the coherence factor 
\begin{equation}
\begin{split}
    \mathds{C}(\tilde{t}) = \left\langle \langle \cos \tilde{\phi} \rangle_{z} \right\rangle
\end{split}
\label{eq:coherence_factor}
\end{equation}
which depends on the spatial average of the  phase difference field $\tilde{\phi}(\tilde{z},\tilde{t})$ between points along the two quasicondensates. The outer brackets indicate an ensemble average which means averaging over many trajectories each sampled from the thermal distribution discussed in Sec.\ \ref{sec:IC1}.  In the Vienna experiments, where one quasicondensate is suddenly split into two, the coherence starts near unity and decays over time as the two quasicondensates decohere \cite{Rauer2018,Pigneur2018}. In the opposite case, where two independent quasicondensates are suddenly coupled, one expects the converse where the coherence starts at zero and grows. This situation has been previously modelled by Horv\'{a}th \textit{et al.} using both the TWA and a truncated conformal space approach \cite{hungarian_paper}. They found that $\mathds{C}(\tilde{t})$ initially grows and then undergoes damped oscillations as it settles down towards a finite constant value. The
coherence factor therefore provides a measure of how the system reaches equilibrium. In this context we note that $\mathds{C}(\tilde{t})$ actually corresponds to an ensemble average of the cosine term in the SG/SG+ Hamiltonian and thus gives information on the exchange of energy between the different parts.  In other words, since the total energy is a constant of the motion,  if the `potential' part of the energy settles down to a constant this suggests the `kinetic' parts of the energy are also constant, at least from an ensemble averaged point of view.  
Our aim in this section is to see if the dynamics of $\mathds{C}(\tilde{t})$ is connected to the caustics. 

\begin{figure}[t]
\begin{tabular}{c}
     {\includegraphics[scale =0.52]{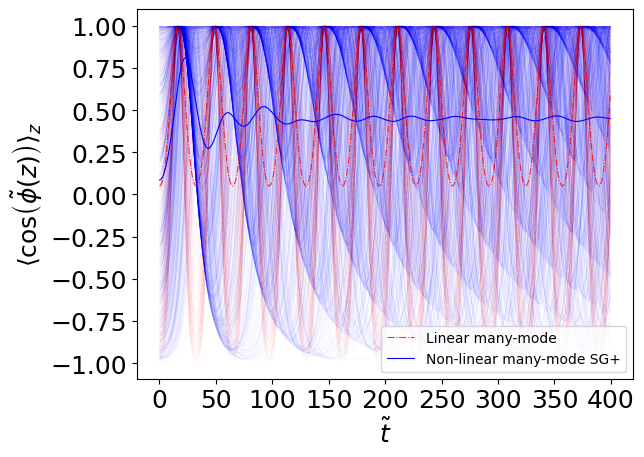}}\\
\end{tabular}
\caption{\label{fig:equilibriation} The two dark lines give the time evolution of the coherence factor $\mathds{C}(\tilde{t})$ defined in Eq.\ (\ref{eq:coherence_factor}) for a linear model (dashed-dotted red) and the SG+ model (solid blue). Both models are multi-mode (many longitudinal modes along $\tilde{z}$) but the SG+ model is nonlinear.  Also included as faint lines are the raw trajectories $\langle \cos \tilde{\phi} \rangle_{z}$ from which $\mathds{C}(\tilde{t})$ is composed. 
As everywhere in this paper, $\langle \ldots \rangle_{z}$ indicates a spatial average.
 This figure highlights that recurrences present in the linear case are suppressed by nonlinearity in the SG+ system. The ensemble average over trajectories with different periods causes $\mathds{C}(\tilde{t})$  to relax towards an equilibrium value in the case of the SG+ model in line with previous experimental observations \cite{Pigneur2018,Rauer2018} and theory  \cite{hungarian_paper}. }
\end{figure}

  In Figure \ref{fig:equilibriation} we plot $\mathds{C}(\tilde{t})$ for two models: the full SG+ model which is many-mode and nonlinear and a linearized version which obeys the equations of motion
\begin{equation}
\begin{split}
    \frac{d \tilde{\phi}}{d \tilde{t}} & = 2 \Gamma \tilde{\rho} - \frac{\Gamma}{2} \frac{\partial^2 \tilde{\rho}}{\partial \tilde{z}^2}\\
    \frac{d \tilde{\rho}}{d \tilde{t}} & = 2 \epsilon \frac{\partial^2 \tilde{\phi}}{\partial \tilde{z}^2} - 2 \mathcal{J} {\tilde{\phi}} .
\end{split}
\label{eq:linear_eom_many_mode_SG_plus}
\end{equation}
This differs from the linearized two-mode approximation defined by Eq.\ (\ref{eq:eomSGplus0Dlinear}) because it describes an elongated multi-mode system. From Figure \ref{fig:equilibriation} we see that $\mathds{C}(\tilde{t})$ for the SG+ model (dark blue curve) does indeed initially grow, undergo damped oscillations and settle down to a non-zero value (the fact that $\mathds{C}(\tilde{t}) \neq 0$ at $\tilde{t}=0$ is due to random fluctuations in the initial conditions: as we include more trajectories we find that the initial value gets smaller). Meanwhile, $\mathds{C}(\tilde{t})$ for the linear model (red dashed curve) executes undamped oscillations and does not settle down to equilibrium. Both models agree during the first oscillation but strongly differ after that.  

It is clear that nonlinearity is important for reaching equilibrium at least as far as global quantities such as $\mathds{C}(\tilde{t})$ are concerned. We can understand this by interpreting the SG+ model as describing a chain of coupled pendula. The nonlinearity of each pendulum means that its period depends on the amplitude of its motion and hence an ensemble of pendula whose motion is initiated together by the quench, but all with different degrees of excitation, will dephase from one another over time so that collective oscillations are damped out. By contrast, linear oscillators have a period independent of their amplitudes of motion and hence remain in phase. 

Apart from the ensemble averages shown by the darker curves in Figure \ref{fig:equilibriation}, we have also included the individual trajectories for $\langle \cos \tilde{\phi} \rangle_{z}$ as fainter curves. The linear model displays harmonic motion and hence perfect revivals whereas the trajectories in the nonlinear model give rise to half-cusp caustics. These caustics overlap in time such that averaging over them causes the coherence to strongly relax after a single period. It is not so much that the caustics cause the relaxation, but rather that both have a common origin in the nonlinearity of the model and hence are generic features of dynamics in complex systems.

\section{Summary and Conclusions}\label{Sec:conclusion}

The sine-Gordon (SG) model is a nonlinear integrable field theory that can be used to describe a wide range of systems from high energy physics to condensed matter physics.  A series of landmark experiments using two coupled 1D atomic quasicondensates \cite{Hofferberth2008,Betz2011,Gring_and_Schmeidmayer_relaxation_paper,Local_emergence_and_correlation_paper,Schweigler_2017,Schmeidmayer_double_light_cone_paper,Rauer2018,Pigneur2018} have realized the SG model in a controllable quantum many body environment. The key parameters can be varied in time allowing the implementation of sudden quenches that excite many modes leading to nonequilibrium dynamics. This is the setting we adopt for the current paper where we use experimentally realistic parameters and compute the dynamics of the number and phase difference fields. However, in contrast  to the usual experimental protocol where the tunnel coupling $J$ is suddenly switched off, we consider quenches where it is suddenly switched on.
While the former case is adapted to studying  dephasing, decay and thermalization between the two subsystems, the many body dynamics is governed by the Tomonaga-Luttinger Hamiltonian describing independent 1D quasicondensates. If instead $J$ is suddenly switched on then the dynamics is that of the full SG model.

Our calculations employ a thermal version of the semiclassical truncated Wigner approximation (TWA) method. More specifically, we propagate a large number of classical field configurations over time with initial conditions sampled from a distribution at thermal equilibrium. The time evolved configurations (trajectories) can be summed to obtain the probability distributions for the observables and we find that these are dominated by singular caustic patterns.  The natural mathematical description of caustics is catastrophe theory that predicts a hierarchy of structurally stable singularities with characteristic shapes that depend on dimension. In two dimensions (e.g.\ number or phase difference versus time) the structurally stable catastrophes are fold lines that meet at cusps.  This is exactly what we find  in both the number and phase differences following a $J$-quench, see Figure \ref{fig:nonlinear1D_dynamics}. The probability distributions develop trains of caustics that are born periodically as cusp points (located at the center of the distribution if there is no tilt) at each plasma period and evolve into pairs of fold lines that gradually move out to the wings where they accumulate. Fold catastrophes manifest as strong non-gaussian fluctuations in the form of inverse square root divergences in the intensity (probability density), as shown in Figure \ref{fig:nonlinear_density_of_trajectories}.

  
 A special case is provided by the dynamics of a two mode system as shown in Figure \ref{fig:0D_dynamics}. Here the equations of motion are the Josephson equations given in Eq.\ (\ref{eq:eomSGplus0D}). The only fluctuations we include in this example are the  quantum fluctuations in the initial relative phase between the two condensates as mandated by the uncertainty principle applied to systems in relative number eigenstates. The two-mode case is relevant to small systems where the higher modes are well above the temperature scale and so any spatial fluctuations are suppressed. By contrast, the many-mode case shown in the other figures includes both quantum fluctuations and thermal fluctuations in the longitudinal modes, i.e.\ thermal occupation of phonon modes in the 1D quasicondensates. Despite the presence of the many longitudinal modes (typically 50 in our calculations, as set by the parameter $N_{L}$) which give rise to highly random looking phase and density profiles as seen in Figure \ref{fig:my-label5}, we find that number and phase caustics survive for experimentally realistic parameters. Furthermore, the qualitative features of the caustics are stable against variations in temperature, quench strength and density imbalance, as seen in Figures \ref{fig:temperature}, \ref{fig:dynamics_no_quench} and \ref{fig:stability}, respectively, and also against the details of the model (in this paper we use the SG+ model which augments the SG model by including longitudinal density gradients). All of these different examples confirm the structural stability of caustics which is the reason why they occur universally without the need for fine tuning. 

The proliferation of caustics over time combined with their migration to the edge of the probability distribution has important consequences for the
long time probability distribution. It takes on the shape of a circus tent featuring a strong central peak due to the cusp tips which are the most singular part of a caustic, flatter intermediate regions, and rapidly decaying edges where the caustics pile up,  see Figure \ref{fig:probability_intensity}. This shape is quite distinct from a gaussian thermal distribution and can be derived assuming an ergodic hypothesis in which individual pendula have equal probability to be anywhere on their energy shell (see Appendix \ref{app:longtime}).
The approach to this equilibrium distribution can be tracked over time using the coherence factor (Figure \ref{fig:equilibriation}) which is a spatial and ensemble average over the phase field and corresponds to the cosine term in the Hamiltonian if the latter is ensemble averaged. The attainment of equilibrium relies on the nonlinearity of the system to dephase itself when ensemble averaged. The caustics also rely on the nonlinearity without which they would reduce to nongeneric perfect revivals (point foci). In this sense caustics are mutually exclusive to recurrences, at least in the statistical sense in which caustics appear in this paper.

Caustics in the SG model could be observed experimentally by measuring the probability density for either the phase difference or the number difference. For example, the phase difference can be obtained by releasing the two quasicondensates from their double well potential and letting them overlap \cite{andrews97,castin97,PethickSmith}. This process must be repeated many times and for as near identical initial conditions and time evolution as possible in order to build up a probability distribution, although due to the structural stability of caustics they will not be particularly sensitive to differences in the experimental setup from run to run. If the probability distribution is obtained for a single time then we expect to see something like that shown in Figure \ref{fig:nonlinear_density_of_trajectories}. In order to observe the time evolution of a caustic, one must then repeat the whole process for a range of different evolution times. This is laborious but technically possible, and since the first cusp caustic appears at half the plasma period the experiment does not need to run for long.


The singular nature of caustics means that they dominate wave fields and are well known in hydrodynamics and optics through phenomena such as tsunamis and gravitational lensing. The results of this paper show that they also occur in the nonequilibrium dynamics of 1D superfluids where a quench plays an analogous role to an underwater earthquake by generating strong excitations beyond the linear regime that are focused in this case by the cosine term in the SG Hamiltonian. The universal properties of catastrophes imply caustics likely also occur in the post-quench dynamics of other condensed matter systems: systems with more degrees of freedom will display higher catastrophes beyond folds and cusps such as hyperbolic and elliptic umbilics \cite{Kirkby2022}.  However, a special feature of the SG model is that it is integrable and so one may ask if that property plays a crucial role in the existence of caustics.  In this context, we note that in classical mechanics caustics are closely associated with the existence of tori in phase space upon which trajectories live \cite{Berry1983Chaos}. Tori are broken up by chaos, and thus caustics are not expected to survive for long in systems which are deep in the chaotic regime. Despite this, the Kolmogorov-Arnold-Moser (KAM) theorem shows that some tori survive in moderately chaotic systems \cite{arnoldbook}, which suggests caustics may also survive in cases where the classical phase-space is mixed, which is the typical case. Indeed, they survive in the three site Bose-Hubbard model \cite{Kirkby2022} which is known to be chaotic \cite{preface_of_many_body_chaos22}. The important problem of extending the KAM theorem to quantum mechanics \cite{glimmers2015} is thus intertwined with the analysis of caustics in quantum systems and provides an interesting direction for extending the present work.

\section*{Acknowledgements} \label{sec:acknowledgements}
    We thank Ryan Plestid for contributions on thermal field sampling in the early stages of this project, Josh Hainge for suggesting the term `circus tent', Isabelle Bouchoule and Maximilian Schemmer for discussions about their 1D Bose gas experiment, and Igor Mazets and Maximilian Pr\"{u}fer for correspondence and advice about the Vienna experiments.  D.O. acknowledges support from the Natural Sciences and Engineering Research Council of Canada (NSERC) through a Discovery Grant (No.\ RGPIN-2017-06605) and A.A. acknowledges a Mitacs Globalink research internship. 
     Research at the Perimeter Institute is supported in part by the Government of Canada, through the Department of Innovation, Science and Economic Development Canada, and by the Province of Ontario, through the Ministry of Colleges and Universities. M.K. would like to acknowledge support from the project 6004-1 of the Indo-French Centre for the Promotion of Advanced Research (IFCPAR), Ramanujan Fellowship (SB/S2/RJN-114/2016), SERB Early Career Research Award (ECR/2018/002085), and SERB Matrics Grant (MTR/2019/001101) from the Science and Engineering Research Board (SERB), Department of Science and Technology (DST), Government of India. M.K. also acknowledges support from the Infosys Foundation International Exchange Program at ICTS and from the Department of Atomic Energy, Government of India, under Project No. 19P1112R\&D.

\appendix
\section{Derivation of the sine-Gordon Hamiltonian}\label{app:sine_gordon_derivation}
 In this appendix we derive the Hamiltonian $H_{\mathrm{SG}}$ as the effective low energy description of two cigar shaped tunnel-coupled quasicondensates \cite{Bouchoule2005,Schweigler_2017} within a classical field description (Gross-Pitaevskii theory). Along the way we also obtain a slightly enhanced Hamiltonian $H_{\mathrm{SG}+}$ that includes contributions from the gradient of density fluctuations that are not included in the sine-Gordon (SG) Hamiltonian. These contributions are not very important for our parameters but play an important conceptual role by introducing an energetic price for a rapidly varying density and hence effectively cut off these fluctuations.  
 
 Assuming tight radial trapping such that each quasicondensate is in its radial ground state, meaning that only longitudinal excitations are taken into account, the second quantized Hamiltonian for the total system be written
\begin{equation}
 \begin{split}
    H & = \int_{-\infty}^{\infty} dz \ \Bigg\{ \sum_{j=1,2} \bigg[- \frac{\hbar^2}{2m}\hat{\psi}_{j}^{\dagger}(z)\frac{\partial^2 \hat{\psi}_j(z)}{\partial z^2} + \\ & U(z)\hat{\psi}_{j}^{\dagger}(z)\hat{\psi}_{j}(z)+\frac{g_{1D}}{2}\hat{\psi}_{j}^{\dagger}(z)\hat{\psi}_{j}^{\dagger}(z)\hat{\psi}_{j}(z)\hat{\psi}_j(z)\bigg] \\
    & -\hbar J \Big[\hat{\psi}_{1}^{\dagger}(z)\hat{\psi}_{2}(z)+\hat{\psi}_{2}^{\dagger}(z)\hat{\psi}_{1}(z) \Big] \Bigg\} \ .
    \label{eq_full_MF_ham}
 \end{split}
\end{equation}
The quantum field operator $\hat{\psi}_j(z)$  annihilates a particle at the point $z$ in the $j^{\mathrm{th}}$ well, where $z$ is the coordinate along the longitudinal direction (long axis of the system).
$m$ is the mass of the particles, $U(z)$ is a possible external potential (in this paper it will be set to zero), $g_{1D}$ controls the interparticle interaction strength, and $J$ is the tunneling frequency between the two wells. In the classical field approximation we replace the field operators by complex functions
\begin{equation}
\hat{\psi}_j(z) \rightarrow \psi_{j}(z) = e^{i\phi_j(z)}\sqrt{n_{1D}+\rho_j(z)} \ .
    \label{eq:quasi_condensate}
\end{equation}
Note that $\phi_{j}$ and $\rho_{j}$ are the phase and density variables for each well rather than their antisymmetric versions  which are used extensively in the main text. 

Let us start by manipulating the kinetic energy term
\begin{align}
-  \sum_{j=1,2}  \int_{-\infty}^{\infty} & dz \, \frac{\hbar^2}{2m}\hat{\psi}_j^{\dagger}(z)\frac{\partial^2 \hat{\psi}_j(z)}{\partial z^2} \\ 
  = \int_{-\infty}^{\infty} dz & \sum_{j=1,2}\frac{\hbar^2}{2m}\bigg[\left(\frac{\partial}{\partial z} e^{-i\phi_j(z)}\sqrt{n_{1D}+\rho_j(z)} \right) \nonumber \\ & \times \left( \frac{\partial}{\partial z} e^{+i\phi_j(z)}\sqrt{n_{1D}+\rho_j(z)}\right) \bigg] \nonumber\\
= \int_{-\infty}^{\infty} dz & \sum_{j=1,2}\frac{\hbar^2}{2m}\bigg(-i\frac{\partial \phi_j}{\partial z} \hat{\psi}_j^{\dagger}+\frac{e^{-i\phi_j}\frac{\partial \rho_j}{\partial z}}{2\sqrt{n_{1D}+\rho_j}}\bigg) \nonumber \\ & \times \bigg(i\frac{\partial \phi_j}{\partial z} \hat{\psi}_j+\frac{e^{i\phi_j}\frac{\partial \rho_j}{\partial z}}{2\sqrt{n_{1D}+\rho_j}}\bigg) \nonumber \\
=\int_{-\infty}^{\infty} dz & \sum_{j=1,2}\frac{\hbar^2}{2m}\bigg\{\hat{\psi}_j^{\dagger}\hat{\psi}_j \left(\frac{\partial\phi_j}{\partial z}\right)^2+\frac{(\frac{\partial \rho_j}{\partial z})^2}{4(n_{1D}+\rho_j)} \nonumber \\ & +i\frac{\frac{\partial \rho_j}{\partial z}\frac{\partial \phi_j}{\partial z}}{2\sqrt{n_{1D}+\rho_j}}[\hat{\psi}_je^{-i\phi_j}-\hat{\psi}_j^{\dagger}e^{i\phi_j}] \bigg\} \nonumber \\
= \int_{-\infty}^{\infty} dz & \sum_{j=1,2}\frac{\hbar^2}{2m}\bigg\{ \hat{\psi}_j^{\dagger}\hat{\psi}_j\left(\frac{\partial\phi_j}{\partial z}\right)^2+\frac{(\frac{\partial \rho_j}{\partial z})^2}{4(n_{1D}+\rho_j)}  \bigg\} \nonumber \\
\approx  \int_{-\infty}^{\infty} dz & \ \frac{\hbar^2}{2m} \Bigg\{ \frac{n_{1D}}{2}\left[ \left(\frac{\partial\phi_s}{\partial z}\right)^2+\left(\frac{\partial\phi_a}{\partial z}\right)^2 \right] \nonumber \\ & +\frac{1}{2 n_{1D}}\left[\left(\frac{\partial\rho_s}{\partial z}\right)^2+\left(\frac{\partial\rho_a}{\partial z}\right)^2 \right] \Bigg\} 
\end{align}
where
\begin{align}
& \phi_a=\phi_1-\phi_2, \quad \phi_s=\phi_1+\phi_2 \\
& \rho_a=\frac{\rho_1-\rho_2}{2},\quad \rho_s=\frac{\rho_1+\rho_2}{2},
\end{align}
and we assume that $n_{1D} \gg \rho_{j}$.
Next we consider the interactions
\begin{align}
    \sum_{j=1,2} & \frac{g_{1D}}{2}\psi_j^{\dagger}\psi_j^{\dagger}\psi_j\psi_j=\sum_{j=1,2}\frac{g_{1D}}{2}[n_{1D}+\rho_j(z)]^2 \nonumber \\
    =&\sum_{j=1,2}\left(\frac{g_{1D}n_{1D}^2}{2}+\frac{g_{1D}\rho_j^2}{2}+g_{1D}n_{1D}\rho_j \right) \nonumber \\
    =&g_{1D}n_{1D}^2+g_{1D}(\rho_s^2+\rho_a^2)+2 g_{1D}n_{1D}\rho_s \ .
\end{align}
Finally, we consider the tunneling term
\begin{align}
    -\hbar & J  \Big[\psi_1^{\dagger}(z)\psi_2(z)+\psi_2^{\dagger}(z)\psi_1(z) \Big] \nonumber \\
    =&-\hbar J \Big[(e^{-i(\phi_1 - \phi_2)}+e^{-i(\phi_2-\phi_1)})\sqrt{n_{1D}+\rho_1} \sqrt{n_{1D}+\rho_2} \Big] \nonumber \\
    =&-2\hbar J\cos(\phi_a)\sqrt{n_{1D}+\rho_1}\sqrt{n_{1D}+\rho_2}\nonumber \\
    =&-2\hbar J\cos(\phi_a)\sqrt{n_{1D}^2+2n_{1D}\rho_s+\rho_s^2-\rho_a^2}\nonumber \\
   \approx & -2\hbar J\cos(\phi_a)(n_{1D}+\rho_s) \approx 
 -2\hbar n_{1D} J\cos(\phi_a) \ .
\end{align}
At very low temperatures the symmetric and antisymmetric components decouple and hence can be treated separately. The lower energy terms are the antisymmetric ones and we obtain the following Hamiltonian
\begin{equation}
\begin{split}
    H_{\mathrm{SG}+}= \int_{-\infty}^{\infty}  dz  \bigg[g_{1D} & \, \rho_a(z)^2 + \frac{\hbar^2 n_{1D}}{4m}\left(\frac{\partial \phi_a}{\partial z}\right)^2 \\ & +\frac{\hbar^2}{4m n_{1D}}\left(\frac{\partial \rho_a}{\partial z}\right)^2 \bigg]  \\
    -  \int_{-\infty}^{\infty} dz & \ 2 \hbar J n_{1D} \cos\left[\phi_a(z)\right] \ . 
\end{split}
\label{eq:fullassym-appendix}
\end{equation}
When the higher wavelength $\rho$ modes are suppressed this reduces to the sine-Gordon model
\begin{equation}
\begin{split}
    H_{\mathrm{SG}}= & \int_{-\infty}^{\infty}  dz \bigg\{g_{1D} \, \rho_a(z)^2 + \frac{\hbar^2 n_{1D}}{4m}\left(\frac{\partial \phi_a}{\partial z}\right)^2 \\ & - 2 \hbar J \, n_{1D} \, \cos\left[\phi_a(z)\right]\bigg\} \ .
\label{eq:sine-gordon-appendix}
\end{split}
\end{equation}

Eq.~\eqref{eq:sine-gordon-appendix} is the finally obtained SG Hamiltonian $H_{\mathrm{SG}}$ which is the low energy description of two cigar shaped tunnel-coupled quasicondensates \cite{Bouchoule2005,Schweigler_2017}.

\section{Derivation of the Tomonaga-Luttinger (TL) Hamiltonian in Fourier space}\label{appendix:hamiltonian initial condition}
In this appendix we derive the Fourier space version of the Tomonaga-Luttinger (TL) Hamiltonian. Starting from Eq.\ (\ref{eq:H_TLplus}), and applying the discrete Fourier decompositions given in Eq.\ (\ref{fourier_modes_convenient_formulation}) and Eq.\ (\ref{eq:Fourier_summation}),  we have

\begin{widetext}
\begin{eqnarray}
    H_{\mathrm{TL+}}(ra)&=& \int_{-\infty}^{\infty}  dz  \, \frac{g_{1D}}{N_L+1} \left(\sum_{k=-N_{L}/2}^{N_L/2}\varrho_{k} e^{i \frac{2\pi k r}{N_L+1}}\right) 
     \times \left(\sum_{l=-N_{L}/2}^{N_L/2}\varrho_{l} e^{i \frac{2\pi l r}{N_L+1}}\right)
    \nonumber \\ &+& \int_{-\infty}^{\infty} dz  \,\frac{\hbar^2 n_{1D}}{4ma^2(N_L+1)}\frac{\partial}{\partial r}\left(\sum_{k=-N_{L}/2}^{N_L/2}\varphi_{k} e^{i \frac{2\pi k r}{N_L+1}}\right)  \times \frac{\partial}{\partial r}\left(\sum_{l=-N_{L}/2}^{N_L/2}\varphi_{l} e^{i \frac{2\pi l r}{N_L+1}}\right) \nonumber \\
    &+& \int_{-\infty}^{\infty} dz  \,\frac{\hbar^2 }{4m n_{1D}a^2(N_L+1)}   \frac{\partial}{\partial r}\left(\sum_{k=-N_{L}/2}^{N_L/2}\varrho_{k} e^{i \frac{2\pi k r}{N_L+1}}\right)  \times \frac{\partial}{\partial r}\left(\sum_{l=-N_{L}/2}^{N_L/2}\varrho_{l} e^{i \frac{2\pi l r}{N_L+1}}\right) \nonumber \\
    &=&a  \sum_{r=-N_L/2}^{N_L/2} \sum_{k=-N_{L}/2}^{N_L/2}\sum_{l=-N_{L}/2}^{N_L/2}\left(\frac{g_{1D}\varrho_k\varrho_le^{i\frac{2\pi(k+l)r}{N_L+1}}}{N_L+1}\right)  \nonumber \\ &-&  a   \sum_{r=-N_L/2}^{N_L/2}   \sum_{k=-N_{L}/2}^{N_L/2}\sum_{l=-N_{L}/2}^{N_L/2}\frac{\hbar^2 n_{1D}}{4ma^2(N_L+1)}  \times \left(\frac{2\pi}{N_L+1}\right)^2kl\varphi_k\varphi_l e^{i\frac{2\pi (k+l)r}{N_L+1}} \nonumber \\
    &-& a  \sum_{r=-N_L/2}^{N_L/2}  \sum_{k=-N_{L}/2}^{N_L/2}\sum_{l=-N_{L}/2}^{N_L/2}\frac{\hbar^2 }{4m n_{1D} a^2(N_L+1)} \times \left(\frac{2\pi}{N_L+1}\right)^2 kl\varrho_k\varrho_l e^{i\frac{2\pi (k+l)r}{N_L+1}}
    \end{eqnarray}
\end{widetext}
where we have split the $z$ coordinate into $N_{L}+1$ grid points separated by distance $a$ so that $z=r\,a$ where $r$ in an integer lying in the range specified by Eq.\ (\ref{eq:numericalgridpoints}). Using the fact that $N_{L}a=L$, and applying the identity $ \sum_{r=-N_L/2}^{N_L/2}e^{i\frac{2\pi(k+l)r}{N_L+1}}=(N_{L}+1)\delta_{k,-l}$ we obtain
     \begin{equation}
     \begin{split}
     H_{\mathrm{TL}+} \approx & a \sum_{k}\sum_{l}g_{1D}\varrho_k\varrho_l\delta_{k,-l} \\ & - a\sum_{k}\sum_{l}\left(\frac{\hbar^2 n_{1D} \pi^2}{mL^2}\right) kl\varphi_k\varphi_l\delta_{k,-l} \\ & - a\sum_{k}\sum_{l}\left(\frac{\hbar^2  \pi^2}{m n_{1D} L^2}\right) kl\varrho_k\varrho_l\delta_{k,-l}
     \end{split}
\label{eq:fourier_app}
   \end{equation}
 where in the second term we have also replaced  $a^2 (N_L+1)^2$ by $L^2$  which holds when $N_{L} \gg 1$. The limits of the summation in Eq.~\eqref{eq:fourier_app} has been omitted for the sake of brevity. We therefore find 
    \begin{equation}
     \begin{split}
      H_{\mathrm{TL}+} \approx \sum_{k} \bigg[ ag_{1D}\varrho_k\varrho_{-k}  + & \frac{a\hbar^2 n_{1D} \pi^2 k^2}{mL^2} \varphi_k\varphi_{-k} \\ & + \frac{a\hbar^2  \pi^2 k^2}{m n_{1D} L^2} \varrho_k\varrho_{-k} \bigg]\\
     =  \sum_{k} \bigg[ ag_{1D}|\varrho_k|^2  + & \frac{a\hbar^2 n_{1D} \pi^2 k^2}{mL^2} |\varphi_k|^2 \\ & + \frac{a\hbar^2  \pi^2 k^2}{m n_{1D} L^2} |\varrho_k|^2 \bigg]
    \label{eq:Pn11}
    \end{split}
    \end{equation}
    where we used the property of real fields that
    \begin{equation}
         \varphi_{-k}=\varphi^{\star}_k, \quad  \mbox{and} \quad \varrho_{-k}=\varrho^{\star}_{k} \ .
    \end{equation}
     Hence the Hamiltonian takes the form given in Eq.~(\ref{Sine-Gordon hamiltonian fourier form plus}) of the main text.

\section{Bench marking of the numerical method}\label{sec:benchmarking}
 The results given in this paper rely on numerically evolving the equations of motion over time for various models [e.g.\ for the full SG+ model the equations of motion are given in Eq.\ (\ref{eq:eomdimensionless})], which we accomplish using the Julia package DifferentialEquations.jl  \cite{rackauckas2017differentialequations}. This implements a Runge-Kutta solver with a user-defined time step. As a measure of the accuracy of our numerical method we use the deviation of the Hamiltonian from its initial value. Since the Hamiltonian should be a constant of motion this gives an indication of the size of the numerical errors. 

\begin{figure}[H]
\begin{tabular}{c}
     {\includegraphics[scale = 0.52]{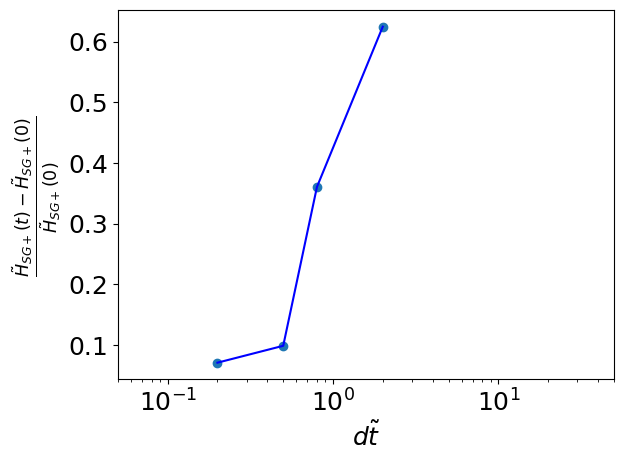}}\\
\end{tabular}
\caption{\label{fig:error_in_time} The relative error in the SG+ Hamiltonian is plotted here as a function of the time step $d\tilde{t}$. The definition of the SG+ Hamiltonian is given in Eq.\ \ref{eq:HSGplus_dimensionless} and should be a constant of the motion were it not for numerical errors. The moving time average of relative error is evaluated after propagating the equations of motion for a total elapsed time of $\tilde{t} = 1000$. All parameter values are the same as in Figure \ref{fig:nonlinear1D_dynamics} including $N_{L} = 50$.}
\end{figure}

\begin{figure}[H]
\begin{tabular}{c}
     {\includegraphics[scale = 0.52]{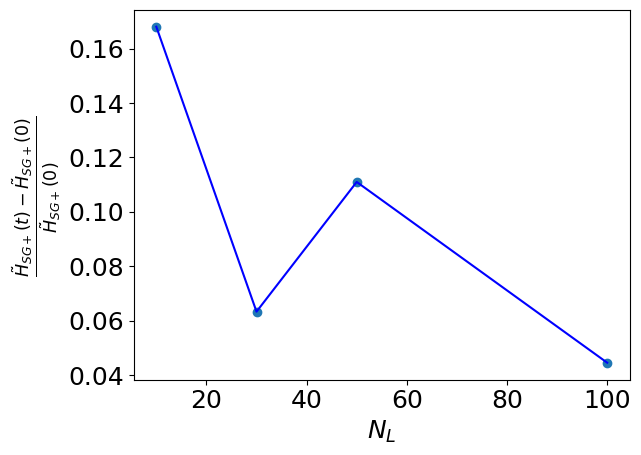}}\\
\end{tabular}
\caption{\label{fig:error_in_space}
The relative error in the SG+ Hamiltonian is plotted here as a function of the number of lattice points $N_{L}$ on the numerical spatial  lattice.  Like in Figure \ref{fig:error_in_time},
the Hamiltonian is evaluated after evolving the equations of motion for a total elapsed time of $\tilde{t} = 1000$. The moving time average of the relative error fluctuates (at around 10\%) but does decrease as d$\tilde{z}$ decreases (or $N_L$ increases). All other parameter values are the same as in Figure \ref{fig:nonlinear1D_dynamics} with $d\tilde{t} = 0.2$}
\end{figure}

In Figures \ref{fig:error_in_time} and \ref{fig:error_in_space} we plot the relative error in the SG+ Hamiltonian given in Eq.\ (\ref{eq:HSGplus_dimensionless}) for different time and spatial resolutions. More precisely, Figure \ref{fig:error_in_time} shows the effect of varying the time step $d\tilde{t}$, whereas Figure \ref{fig:error_in_space} shows the effect of varying the number of grid points $N_{L}$ which sets the spatial step $d\tilde{z}$.  In both cases we have evolved the system for a total elapsed time of $\tilde{t} = 1000$ which corresponds to the longest times we use in this paper (for the calculation of the long-term distribution shown in Figure \ref{fig:probability_intensity}), and also
taken an ensemble average over 100 different trajectories similar to those in Figure \ref{fig:nonlinear1D_dynamics}.  Furthermore, we also performed a moving time average of 30-time steps around $\tilde{t} = 1000$ to average out the effect of fast oscillations. 

 As expected, the relative error decreases as $d\tilde{t}$ and $d\tilde{z}$ decrease. For all the calculations in the main part of this paper we chose 
 $d\tilde{t} = 0.2$ and $N_L = 50$ because this keeps the relative error below 10\% and does not significantly slow down the simulations.

\section{Caustic curve}
\label{app:envelope}

In this appendix we use the exact solution for the motion of a pendulum to calculate the caustic curve plotted as the solid black line in Figure \ref{fig:0D_dynamics}. The caustic is in fact the envelope of a whole family of trajectories. To begin, we take the equations of motion for the SG model given in Eq.\ (\ref{eq:eomdimensionless}) and drop the second order derivative term proportional to $\epsilon$ which couples the different pendula. Next, we make the change of variables 
\begin{equation}
    \tilde{t}= A t, \quad  \tilde{\rho} = B p, \quad \tilde{\phi}= 2y
    \label{eq:pendulumvariables}
\end{equation}
where
\begin{equation}
A=\frac{1}{2}\frac{1}{\sqrt{\mathcal{J} \Gamma}} \quad , \quad B=2 \sqrt{\frac{\mathcal{J}}{\Gamma}} 
\end{equation}
so the equations of motion simplify to
\begin{eqnarray}
\frac{dy}{dt} & = & p  \label{eq:dydt} \\
\frac{dp}{dt} & = & - \frac{1}{2} \sin 2 y  \ .
\label{eq:dpdt}
\end{eqnarray}
These equations are Hamilton's equations  obtained from a standard pendulum hamiltonian of the form
\begin{equation}
    H(y,p)= \frac{p^2}{2}+\frac{1}{2} \sin^2 y \ .
    \label{eq:pendulumHamiltonian}
\end{equation}  

The equations of motion given in Eqns.\ (\ref{eq:dydt}) and (\ref{eq:dpdt}) have exact solutions in terms of the Jacobi elliptic functions $\mathrm{sn}[u\vert m]$ and $\mathrm{cn}[u \vert m]$ \cite{NISTHandbook}. For the case relevant to us where the pendulum starts at angle $y_{0}$, with zero initial angular momentum, they are  
\begin{eqnarray}
y(t,y_{0}) & = & \arcsin \{ \sin y_{0} \
\mathrm{sn}[t + K(\sin y_{0}) \vert \sin y_{0} ] \} \label{eq:pendulumsoly} \\
p(t,y_{0}) & = & \sin(y_{0}) \ \mathrm{cn}[t + K(\sin y_{0}) \vert \sin y_{0}] \label{eq:pendulumsolp}
\end{eqnarray}
where $K(m)=\int_{0}^{\pi/2} d \theta /\sqrt{1-m^2 \sin^2 \theta} $ is the complete elliptic integral of the first kind \cite{NISTHandbook} (we caution the reader that some computer packages such as \textit{Mathematica} use the syntax $K(m^2)$ for this integral).

Caustics occur when trajectories are focused, in other words they are the places where the trajectory does not change (to first order) when the initial conditions are varied. Thus, caustics in the momentum variable $p$ occur when $dp/dy_{0}=0$ since the initial condition here is specified by $y_{0}$. 
 By differentiating Eq.\ (\ref{eq:pendulumsolp}) an implicit expression for the position of the caustics can be found \cite{odell99} 
\begin{eqnarray}
&&  \mathrm{sn}(u|m)\mathrm{dn}(u|m) \left(\frac{\mathrm{E}(\mathrm{am} (-t \vert m)\,\vert m)}{\cos (y_{0})}+t \cos (y_{0})\right) \nonumber  \\ && -\cos (y_{0}) \mathrm{cn}(u \vert m)=0  \label{eq:caustic-condition}
\end{eqnarray}
where $u=t+K(\sin y_{0})$, $m=\sin y_{0}$, $\mathrm{E}(u \vert m)$ is an elliptic integral of the second kind, $\mathrm{dn}(u \vert m)$ is another Jacobi elliptic function,  and $\mathrm{am}(u \vert m)=\arcsin [\sin (\phi)/m]$ is the Jacobi amplitude \cite{NISTHandbook}. Finding the roots $y_{0}$ of Eq.\ (\ref{eq:caustic-condition}) numerically at each value of the time gives pairs of values $(y_{0},t)$ that can then be put back into Eq.\ (\ref{eq:pendulumsolp}) to yield the black curve for the caustic shown in Figure \ref{fig:0D_dynamics}. The match to the numerics is very good.

\section{Derivation of ergodic (``circus tent'') probability distribution at long times}\label{app:longtime}

In this appendix we outline the derivation of an analytic approximation to the probability distribution for the number difference at long times, as shown in Figure \ref{fig:probability_intensity}. This derivation is based upon a calculation given in Ref.~\onlinecite{ergo_paper} and assumes that the average behaviour of a continuous chain of coupled pendula (the mechanical system that underlies the  sine-Gordon model) can be described by a suitably `ergodized' single pendulum. 

To keep the calculation general we use the pendulum Hamiltonian in standard form as given in Eq.\ (\ref{eq:pendulumHamiltonian}).
With this hamiltonian we define a microcanonical probability density in phase space: 
\begin{equation}
d_{m}(y,p;y_{0}) = \frac{\delta[H(y,p)-H(y_{0},p)]}{\int \int dy \ dp \ \delta[H(y,p)-H(y_{0},p)]}  
\label{eq:microcanonical}
\end{equation}    
 where $y_{0}$ is the initial angle of the pendulum which fixes its total energy to be $E=(1/2) \sin^{2} y_{0}$ if the the initial angular momentum is zero (this is the appropriate initial condition for the tunneling quench considered in this paper where the initial number difference is taken to be zero), and the denominator ensures that $d_{m}$ is normalized to unity.  A microcanonical distribution has equal probability to be anywhere on its energy shell (in this case a closed curve in $y,p$ phase space) and thus by adopting Eq.\ (\ref{eq:microcanonical}) we are making an \textit{ergodic hypothesis}. This does not hold for a single pendulum starting at position $y_{0}$ since it will spend the most time at its turning points $y=\pm y_{0}$, but when averaged over $y_{0}$ and $y$ (see below) it gives a very good approximation at long times, as can be seen in Figure \ref{fig:probability_intensity}.

The normalization integral can be evaluated exactly by re-expressing the delta function using the relation $\delta[g(x)]=\sum_{i} \delta(x-x_{i})/\vert g'(x_{i}) \vert$, where $x_{i}$ are the roots of $g(x)$. In the present case this gives
\begin{equation}
\begin{split}
    \delta[(p^2+\sin^2 y - \sin^2 y_{0})/2]= & \frac{\delta[p-p_{1}]}{\vert p_{1}\vert}+ \frac{\delta[p-p_{2}]}{\vert p_{2}\vert} \\ = & 2 \frac{\delta[p-p_{1}]}{\vert p_{1}\vert}
    \end{split}
\end{equation}
where $\vert p_{1} \vert = \vert p_{2} \vert =\sqrt{\sin^2 y_{0} - \sin^2 y }$. In obtaining this expression we have used the fact that for values of $y$ within the range accessed by the pendulum, there are two values of $p$ where the integral crosses the energy shell. The integral over $p$ is now trivial due to the delta function and the integral over $y$ can be performed by putting $\sin y = \sin y_{0} \sin \zeta$ so that
\begin{widetext}
\begin{equation}
\begin{split}
\label{eq:ellipticintegraltype}
  2 \int_{-y_{0}}^{y_{0}} \frac{dy}{\vert p(y,y_{0}) \vert} =   2 \int_{-y_{0}}^{y_{0}} \frac{dy}{\sqrt{\sin^{2}y_{0}-\sin^2 y}}   = 2 \int_{-\pi/2}^{\pi/2} \frac{d \zeta}{\sqrt{1- \sin^2 y_{0} \sin^2 \zeta}}   = 4 \int_{0}^{\pi/2} \frac{d \zeta}{\sqrt{1- \sin^2 y_{0} \sin^2 \zeta}}   = 4 K(\sin y_{0}) \ .
  \end{split}
\end{equation}
\end{widetext}
Therefore, the normalized microcanonical probability density can be written as
\begin{equation}
\begin{split}
d_m(y,p;y_{0})= & \frac{1}{4 K(\sin y_{0})}  \delta[(p^2+\sin^2 y - \sin^2 y_{0})/2] \\
= & \frac{1}{2 K(\sin y_{0})}  \delta(p^2+\sin^2 y - \sin^2 y_{0})
\end{split}
\end{equation}
where we have used the property of delta functions that $\delta(\alpha x)= (1/\alpha)\delta (x)$.

The initial condition for our dynamics is such that the number difference is well defined but the phase difference is completely undefined.  We must therefore average the microcanonical probability density over all $y_{0}$. This gives the phase space probability density relevant to $J$-quenches as being
\begin{equation}
    W(y,p)= \frac{1}{\pi} \int_{- \pi/2}^{\pi/2} dy_{0} \ d_{m}(y,p;y_{0})
\end{equation}
where we employ the notation $W$ to indicate that this is a classical version of the Wigner function.
 The properties of the delta function can once more be used to write 
\begin{equation}
\delta(p^2+\sin^2 y - \sin^2 y_{0}) = \sum_{i} \frac{\delta(y_{0}-y_{0i}) \theta(\cos y - \vert p \vert)}{2\sqrt{p^2+\sin^2 y}\sqrt{\cos^2 y -p^2}}
\end{equation}
where $\theta(x)$ is the Heaviside step function. The integral over $y_{0}$ can now be evaluated exactly to give
\begin{equation}
    W(y,p)= \frac{2}{4 \pi} \frac{ \theta(\cos y - \vert p \vert)}{K(\sqrt{p^2+\sin^2 y})\sqrt{p^2+\sin^2 y}\sqrt{\cos^2 y -p^2}} .
\end{equation}

\begin{figure}
\includegraphics[scale=0.55]{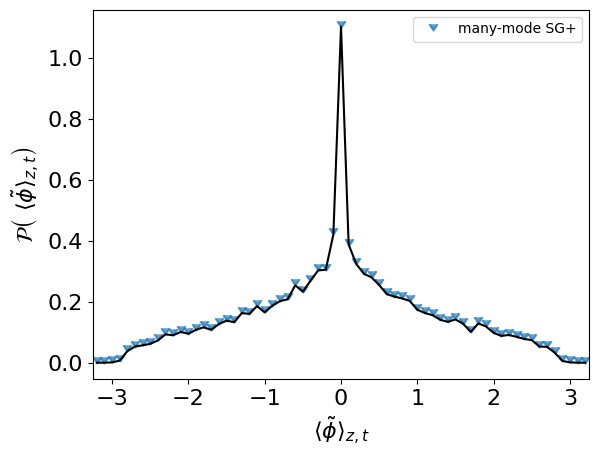}
\caption{Long-time probability distribution for the phase difference $\tilde{\phi}$ . The data points are from the SG model averaged over the spatial coordinate $z$ and also over a time window ranging from $\tilde{t} = 800$ to $\tilde{t} = 980$ to remove fluctuations. The solid line is not a fit or theoretical curve but passes through all the data points to help guide the eye..}
\label{fig:long_time_phase}
\end{figure}

The final step is to integrate out the $y$ coordinate to obtain the probability distribution $\mathcal{P}_{\mathrm{CT}}(p)$ for $p$ alone
\begin{equation}
    \mathcal{P}_{\mathrm{CT}}(p)=\int_{-\pi/2}^{\pi/2} dy \ W(y,p) \ ,
\end{equation}
where ``CT'' stands for circus tent.
Although this integral cannot be done analytically, it can be put in a   form which is convenient to evaluate numerically. Denoting $m=\sin y_{0}=\sqrt{p^2+\sin^2y}$, one finds that 
\begin{widetext}
\begin{equation}
\label{eq:circustent}
   \mathcal{P}_{\mathrm{CT}}(\tilde{\rho})=\frac{1}{2 \pi B} \int_{\tilde{\rho}^2/B^2}^{1} \frac{dm}{K(m)\sqrt{m(1-m)(m-\tilde{\rho}^2/B^2)(1+\tilde{\rho}^2/B^2-m)}} 
\end{equation}
\end{widetext}
where we have also converted back from angular momentum $p$ to number difference $\tilde{\rho}$ using Eq.\ (\ref{eq:pendulumvariables}).
This equation is given in the main text as Eq.\ (\ref{eq:circustenta}) and is plotted in Figure \ref{fig:probability_intensity} where it is  compared against the long-time spatially and temporally averaged numerical data for the various nonlinear models considered in this paper.
As can be seen in Figure \ref{fig:probability_intensity},  $\mathcal{P}_{\mathrm{CT}}$ is characterized by a diverging (yet normalizable) peak at the center and then relatively flat wings until it drops sharply to zero at the edges.  In Ref.~\onlinecite{ergo_paper} it is shown that $\mathcal{P}_{\mathrm{CT}}(\tilde{\rho})$ diverges logarithmically at the origin $\tilde{\rho}=0$ and also tends suddenly to zero with logarithmic singularities at $\tilde{\rho}=\pm B$. Both these non-thermal features can be attributed to the presence of caustics.
We have also numerically calculated the long-time probability distribution for the relative phase variable, and it is plotted in Fig.\ \ref{fig:long_time_phase}. Like Fig.\ \ref{fig:probability_intensity}, it has a nonthermal circus tent-like shape with a peak at the center, relatively flat wings and then sharply drops to zero at the edges.

\section{Pendulum at thermal equilibrium}\label{app:thermalpendulum}

In Figure \ref{fig:probability_intensity} the long time probability distribution for the number difference is compared against the ergodic prediction derived in Appendix \ref{app:longtime}, and also against the thermal equilibrium prediction. In this Appendix we explain how to calculate the latter case. 
In order to make the calculation tractable we make the assumption that the SG+ model can be approximated by a thermal ensemble of independent pendula. We also adopt the same notation as Appendix 
\ref{app:longtime} and hence work with a pendulum Hamiltonian in the standard form $H=(1/2)(p^2 + \sin^{2} y)$. This is related to the two mode Hamiltonian $H_{2M}=\Gamma \tilde{\rho}^2-2 \mathcal{J} \cos \phi$ by $H=H_{2M}/8 \mathcal{J}+1/4$.

We proceed in two steps: we first calculate the probability distribution $\mathcal{P}_{E}(p)$ for the momentum variable $p$ (that here plays the role of the number difference) for a fixed energy $E$. Secondly, we assume our system is at thermal equilibrium with a bath at temperature $T$ such that the relative probability of any energy is given by the Boltzmann factor $\exp[-E/ T]$. Thus the thermal probability distribution is
\begin{equation}
\label{eq:totalprob}
    \mathcal{P}_{T}(p) = \frac{1}{Z} \int_{0}^{\infty} \mathcal{P}_{E}(p) \ e^{-E/T} D(E) \ dE
\end{equation}
where $Z$ is a normalizing factor (found numerically) and $D(E)$ is the density of states.

The probability distribution $\mathcal{P}_{E}(p)$ at fixed $E$ is proportional to $1/\dot{p}$ as this determines how long the pendulum spends at each value of $p$. According to Hamilton's equation $\dot{p}=-\partial H / \partial x = -(1/2)\sin 2 y$, and using the fact that $\sin y = \sqrt{2E-p^2}$, we find that this probability distribution for a fixed value of $E$ is
\begin{equation}
\label{eq:P_E}
\mathcal{P}_{E}(p)= \frac{\mathcal{N}}{(1/2)\sin (2 \arcsin{\sqrt{2E-p^2}})}  ,
\end{equation}
where $\mathcal{N}$ is a normalization factor given by the period of the motion. Two cases must be distinguished: for $E<1/2$ the energy is less than the separatrix and the  pendulum undergoes vibrational motion (also known as librational motion in some literature). Conversely, when $E>1/2$ the energy is above the separatrix and the pendulum undergoes rotational motion.

For motion below the separatrix we have $\vert p \vert < p_{\mathrm{max}} = \sqrt{2E}$. We must therefore supplement the expression for $\mathcal{P}_{E}(p)$ with the condition that it is zero if $\vert p \vert > p_{\mathrm{max}}$ and this ensures that $\mathcal{P}_{E}(p)$ is real. 
 $\mathcal{N}$ is given in this case by
\begin{equation}
\mathcal{N}=\frac{1}{2 \ K(\sqrt{2 E})}    
\end{equation}    
where, as in Appendix \ref{app:longtime}, $K$ is the complete elliptic integral of the first kind. 

For motion above the separatrix we have $\sqrt{2E-1} < \vert p \vert < \sqrt{2E}$ and $\mathcal{P}_{E}(p)$ is zero outside this range. 
 $\mathcal{N}$ is now given by
\begin{equation}
\mathcal{N}=\frac{\sqrt{2E}}{4 \ K(1/\sqrt{2E})} .   
\end{equation} 

To obtain the total thermal probability distribution $\mathcal{P}_{T}(p)$ given in Eq.\ (\ref{eq:totalprob}) we need the  density of states $D(E) \equiv dn/dE$, where $n$ is the number of states  below energy $E$. According to the Bohr-Sommerfeld rule $n=S(E)/(2 \pi \hbar)$, where the action $S(E)=\oint p \, dy$ is the area in phase space enclosed by the energy contour $E$. However, assuming that our Hamiltonian $H$ is in units $\hbar \omega$ then the $2 \pi \hbar$ factor is absorbed into the definitions of $p$ and $y$ and we have $D(E)=(d/dE)\oint p \, dy$. Below the separatrix we have
\begin{equation}
\oint p(y)dy=4 \int_{0}^{\arcsin\sqrt{2E}}\sqrt{2E- \sin^2 y} \ dy    
\end{equation}
and putting $2E=\sin^2 y_{0}$ we find
\begin{eqnarray}
D_{<}(E)= & \frac{d}{d E} & \oint p(y)dy \nonumber \\ &= & 4 \int_{0}^{\arcsin \sqrt{2E}}\frac{dy}{\sqrt{\sin^2 y_{0}- \sin^2 y} } \nonumber \\
& = & 4 K(\sqrt{2E})
\label{eq:dosbelow}
\end{eqnarray}
where the integral is performed in a similar fashion to the one in Eq.\ (\ref{eq:ellipticintegraltype}) and the subscript ``$<$'' indicates that this is the expression valid below the separatrix. Above the separatrix we find that the area enclosed in phase space between two oppositely rotating states of the same energy is
\begin{equation}
\oint p(y)dy= 2 \int_{-\pi/2}^{\pi/2}\sqrt{2E- \sin^2 y} \ dy    
\end{equation}
and thus
\begin{eqnarray}
D_{>}(E) = & \frac{d}{d E} & \oint p(y)dy \nonumber \\ & = &  2 \int_{-\pi/2}^{\pi/2}\frac{dy}{\sqrt{2E- \sin^2 y} } \nonumber \\
& = & \frac{4}{\sqrt{2E}} K\left(\frac{1}{\sqrt{2E}} \right) \ .
\label{eq:dosabove}
\end{eqnarray}
Due to the fact that above the separatrix $2E > \sin^2 y$ we no longer need to make the substitutions  $2E=\sin^2 y_{0}$ and $\sin y = \sin y_{0} \sin \zeta$, and the integral is straightforward. The subscript ``$>$'' indicates that this expression holds above the separatrix.

We now have all the necessary ingredients to perform the integral for $\mathcal{P}_{T}(p)$ which we do numerically.
The two contributions, one from below the separatrix and one from above, are added together to get the total. Interestingly, both density of states factors, Eqns.\ (\ref{eq:dosbelow}) and (\ref{eq:dosabove}), diverge at the separatrix such that the two contributions individually display singular features but remarkably these cancel out when the two parts are added and result in the smooth gaussian curve plotted in Figure \ref{fig:probability_intensity}.

In order to compare the thermal distribution against the quenched (followed by integrable SG evolution) distribution derived in Appendix \ref{app:longtime} we need to choose a temperature $T$ for the thermal distribution $\mathcal{P}_{T}$. We do this by matching the expectation value of the energy $\langle E \rangle$ for both distributions. In the quenched case the initial state corresponds to an ensemble of pendula with different starting angles $y_{0}$ and zero kinetic energy. Each starting angle  in the range $-\pi/2 < y_{0} \le \pi/2$ is equally probable in our $J$-quench. Therefore
\begin{equation}
\label{eq:Eexpectationquench}
    \langle E \rangle_{\mathrm{quench}}=\frac{1}{\pi} \int_{-\pi/2}^{\pi/2} \frac{1}{2} \sin^2 y_{0} \  dy_{0}= \frac{1}{4} \ .
\end{equation}
To calculate $\langle E \rangle$ in the thermal case we compute 
\begin{equation}
   \langle E \rangle_{T}= \frac{1}{\zeta} \int_{0}^{\infty} E \, e^{-E/T} \, D(E) \, dE
\end{equation}
numerically for a large number of different values of $T$, performing the integrals below and above the separatrix separately and adding the results. Here $\zeta=\int_{0}^{\infty} e^{-E/T} \, D(E) \, dE$ gives the normalization factor.
We then fit a curve to the results and find the value of $T$ that best matches the  result given in Eq.\ (\ref{eq:Eexpectationquench}). We find that $T=0.184$ gives the best match. Putting back the units this result is 
\begin{equation}
    \frac{k_{B}T}{8 \mathcal{J} \hbar c / \xi_{h}} = \frac{k_{B}T}{16   J  \hbar  K/\pi}=0.184
\end{equation}
where $c$ is the speed of sound and $K$ is the Luttinger parameter and $J$ is the tunnel coupling rate between the two wells. In this paper we take $K=25$ and $J=30$ Hz (see Table \ref{parameter_table}) giving a temperature in SI units of 5.4 nK.





%

\end{document}